%% file: main.tex
\documentclass[a4paper,fleqn,usenatbib]{mnras}

\usepackage{mathptmx}
\usepackage[T1]{fontenc}
\usepackage{ae,aecompl}

\usepackage{graphicx}	% Including figure files
\usepackage{amsmath}	% Advanced maths commands
\usepackage{amssymb}	% Extra maths symbols

\usepackage{color}
\usepackage{soul}
\usepackage{ulem}
\usepackage{subcaption}  % make subfigures
\newcommand{\unitg}{\text{cm s$^{-2}$}} % unit of acceleration
\newcommand{\unity}{\text{g cm$^{-2}$}} % unit of column depth
\newcommand{\unitrho}{\text{g cm$^{-3}$}} % unit of density
\newcommand{\difsub}[1]{ \mathrm{d}_{#1} } % total derivative shortcut
\newcommand{\parsub}[1]{ \partial_{#1} } % partial derivative

\newcommand{\msol}{M_{\odot}}
\newcommand{\cross}{\times}
\newcommand{\bvec}[1]{\pmb{#1}} % bold vector in equations
\newcommand{\bruntvaisala}{Brunt-V\"{a}is\"{a}l\"{a} }

\newcommand{\ign}[1]{{#1}_{\text{ign}}}
\newcommand{\yign}{y_{\text{ign}}}

\newcommand{\Tcrust}{T_{\text{crust}}}
\newcommand{\Tpeak}{T_{\text{peak}}}

\newcommand{\alphaa}{0.23}
\newcommand{\alphab}{0.25}
\newcommand{\yigna}{2 \times 10^{11}}
\newcommand{\yignb}{6 \times 10^{11}}
\newcommand{\Tcrusta}{3 \times 10^{8}}
\newcommand{\Tcrustb}{5 \times 10^{8}}
\newcommand{\Tpeaka}{4.8 \times 10^{9}}
\newcommand{\Tpeakb}{5.1 \times 10^{9}}

\title[Superburst Oscillations]{Superburst oscillations: ocean and crustal modes excited by Carbon-triggered Type I X-ray bursts}

\author[Chambers et al.]{
F.R.N. Chambers,$^{1}$\thanks{E-mail: frnchambers@uva.nl}
A.L. Watts,$^{1}$
Y. Cavecchi,$^{2,3}$
F. Garcia,$^{1,4}$
L. Keek$^{5}$
\\
$^{1}$Anton Pannekoek Institute for Astronomy, University of Amsterdam, Postbus 94249, 1090 GE Amsterdam, The Netherlands\\
$^2$ Department of Astrophysical Sciences, Princeton University, Peyton Hall, Princeton, NJ 08544, USA\\
$^3$ Mathematical Sciences and STAG Research Centre, University of Southampton, SO17 1BJ, UK\\
$^4$ Helmholtz-Zentrum Dresden-Rossendorf, POB 51 01 19, 01314 Dresden, Germany\\
$^5$ Department of Astronomy, University of Maryland, College Park, MD 20742, USA
}

\date{6 April 2018}

\pubyear{2018}

\begin{document}
\label{firstpage}
\pagerange{\pageref{firstpage}--\pageref{lastpage}}
\maketitle

% This is a simple template for authors to write new MNRAS papers.
% The abstract should briefly describe the aims, methods, and main results of the paper.
% It should be a single paragraph not more than 250 words (200 words for Letters).
% No references should appear in the abstract.
\begin{abstract}

Accreting neutron stars (NS) can exhibit high frequency modulations in their lightcurves during thermonuclear X-ray bursts, known as burst oscillations. The frequencies can be offset from the spin frequency of the NS by several Hz, and can drift by 1-3 Hz. One possible explanation is a mode in the bursting ocean, the frequency of which would decrease (in the rotating frame) as the burst cools, hence explaining the drifts. Most burst oscillations have been observed during H/He triggered bursts, however there has been one observation of oscillations during a superburst; hours' long Type I X-ray bursts caused by unstable carbon burning deeper in the ocean. This paper calculates the frequency evolution of an oceanic r-mode during a superburst. The rotating frame frequency varies during the burst from $4-14$ Hz, and is sensitive to the background parameters, in particular the temperature of the ocean and ignition depth. This calculation is compared to the superburst oscillations observed on 4U-1636-536. The predicted mode frequencies ($\sim 10$ Hz) would require a spin frequency of $\sim 592$ Hz to match observations; 6 Hz higher than the spin inferred from an oceanic r-mode model for the H/He triggered burst oscillations. This model also over-predicts the frequency drift during the superburst by $90 \%$.

\end{abstract}


%%%%%%%%%%%%%%%%%%%%%%%%%%%%%%%%%%%%%%%%%%%%%%%%%%
%%%%%%%%%%%%%%%%%% BODY OF PAPER %%%%%%%%%%%%%%%%%

\section{Introduction}\label{section:introduction}
Buoyant modes in rotating oceans are a topic in fluid dynamics with a long history and a wide range of application \citep{Pedlosky87}. Interesting in and of their own right, they also have the potential to explain phenomena that have puzzled neutron star astronomers for nearly two decades. 

Accreting neutron stars (NS) develop thin surface oceans of hydrogen, helium and heavier elements. Heat is generated by accretion and thermonuclear burning which can be unstable and explosive, resulting in Type I (thermonuclear) X-ray bursts. Most of these bursts are caused by unstable H/He burning \citep[for a review, see][]{Bildsten98b}, and last seconds to minutes.  Timing analysis of these bursts reveals periodic frequencies which are either at, or very close to, the NS spin frequency \citep[known independently for some stars; for a review see][]{Watts12}. These {\it burst oscillations} \citep{Strohmayer96b} must arise from some kind of surface brightness pattern; however the mechanism responsible for these oscillations has yet to be identified.

One possibility is the development of ocean modes. These modes can give rise to large scale patterns, and the variety of different families (driven by different restoring forces) can lead to a range of observed frequencies.
Moreover, ocean modes could plausibly be excited by bursts which is why \cite{Heyl04} suggested them as a potential explanation for burst oscillations. Heyl used some simple arguments, based on observed properties, to constrain the class of modes that could potentially fit the data.  The key constraints were:
(1) the rotating frame frequency of the mode should be  $\sim 1$ Hz, and the azimuthal eigenvalue, $m$, should be small, making the observed frequency close to the spin frequency;
(2) since observed frequencies drift upwards to the spin frequency (when known independently) as the layer cools, the modes should travel in the opposite sense to the star's rotation (retrograde);
(3) the modes should have no latitudinal nodes, maximising visibility since modes are squeezed near the equator;
(4) modes should have no radial nodes which would separate modes (in frequency) with a similar angular dependence, helping to ensure that only a single mode is excited during a burst.

Buoyant r-modes, with very low $m$, are the most promising candidate to fit these constraints.
These modes are strongly influenced by the Coriolis force, and hence do not occur on non-rotating stars. Buoyancy acts as the restoring force in the hot bursting ocean due to large temperature and compositional gradients.
This buoyancy could also give rise to g-modes or Kelvin modes. These are discounted since, in the case of the g-modes, perturbations would be squeezed into a narrower band near the equator \citep{Heyl04} (where obscuration by the accretion disk would also be more of an issue) and therefore less visible than other candidates. In the case of the Kelvin modes, these are prograde which implies a negative drift in the aftermath of a burst.

One problem with the buoyant r-mode model is that the amount of frequency drift predicted, as the burning layer cools, is larger (at $\sim 10$ Hz) than that observed (which is $\sim$ 1-3 Hz).  Work by \cite{Piro05b} (hereafter PB05b) explored the idea that the ocean mode might transition into a crustal interface wave, thereby curtailing the drift. This particular idea has since been ruled out by \citet{Berkhout08} due to the weak coupling between the ocean mode and the crustal interface wave. The drift size, therefore, remains an issue.

Nevertheless an ocean mode of some form remains a reasonable candidate to explain the burst oscillation phenomenon, since it provides a natural way to explain frequency drifts.
Other mechanisms that may be involved in this phenomenon include: a change in spin frequency of the surface layers due to expansion \citep{Cumming02}; a possible flame spread model \citep{Cavecchi16}; and a cooling wake model \citep{Mahmoodifar16}. These models have problems; the frequency drift in the expansion model under-predicts what should be observed; the flame spread model can account for oscillations in the rise, but not the tail, where most drifts are measured, and requires a higher magnetic field than expected for most sources; and the cooling wake model requires enhanced cooling from an as yet unidentified physical mechanism. It is also possible that convection in the ocean of the NS must be taken into account \citep{Medin15,Garcia18}. More extensive studies, however, need to be performed. Magnetic modes, too, could play a role \citep{Heng09}.

Another class of Type I X-ray bursts are {\it superbursts}; hours' long explosions caused by unstable C burning that takes place much deeper in the ocean than H/He burning. They are quite different to H/He bursts, being $10^2-10^3$ times more energetic and with longer recurrence times of approximately $1$ yr as compared to hours or days \citep[see][for reviews]{Kuulkers04,Cumming04b,Strohmayer06,intZand17}. 
A superburst on 4U 1636-536 is the only one to have been observed in high time resolution. In the period from 10$^2$ secs to 10$^{3}$ secs after the start of the superburst it exhibited burst oscillations \citep{Strohmayer02a}. 
This source has also exhibited oscillations during H/He X-ray bursts \citep{Strohmayer98a,Muno02b}, providing the opportunity to compare oscillations in H/He bursts and superbursts, for the same star.
Oscillations during the superburst were more stable and at a slightly higher frequency than their H/He burst counterparts. The small frequency drift of the superburst oscillations was consistent with orbital Doppler shifts, while the H/He burst oscillations drift by rather more. The amplitude of the oscillations during the superburst was weaker than that of the H/He burst. There have also been weak higher frequency oscillations observed during this superburst \citep{Strohmayer14}.

Since buoyant r-modes in H/He bursts are thought to develop due to the temperature and compositional gradients present after the burst ignites, it is plausible that such modes should also develop in the aftermath of superbursts when strong temperature and composition gradients are also present. The only mode calculations carried out to date, however, have used simple models for the composition of burst ashes at depths appropriate for H/He bursts (PB05b). Buoyant r-mode calculations have not yet been carried out for depths and compositions appropriate for the aftermath of Carbon burning. 

We follow similar analysis to PB05b, calculating the frequency and frequency-drift of a buoyant r-mode. This paper, though, calculates these quantities for a burst deeper in the ocean in the aftermath of a superburst. This study aims to find stable modes on the cooling background and to test the sensitivity of mode frequencies to differences in the cooling profile.
We start, in Section~\ref{section:perturbations}, by outlining the perturbation equations that are solved to obtain the mode frequencies. Section~\ref{section:background} describes the thermal evolution of the burning layer during the burst tail, since this is the cooling background in which the modes form. Section~\ref{section:results} discusses the solutions that are found. In Section~\ref{section:conclusion} we discuss the implications for models of the burst oscillation mechanism.

\section{Perturbation Equations} \label{section:perturbations}
The outer layers of an accreting NSs are made of an accreted H and He ocean which burns to heavier elements as material moves deeper into the star. This ocean extends down as far as the outer crust; a lattice of heavy ions \citep{Chamel08} bound by Coulomb forces surrounded by a sea of degenerate electrons. The ocean/crust transition is usually defined using the Coulomb coupling parameter which is the ratio of Coulomb energy to thermal energy;
\begin{equation}
  \label{eq:gamma}
  \Gamma = \frac{(Ze)^2 / a}{k_B T} ,
\end{equation}
where $Z$ is the charge per ion in units of $e$, electron charge; $a = \left( 4 \pi n_{i} / 3 \right)^{-1/3}$ is the average ion spacing, with $n_{i}$ the ion number density; $T$ is the temperature; and $k_B$ the Boltzmann constant. The ocean/crust transition occurs at $\Gamma \approx 173$ \citep{Farouki93}, giving a transition density around  $10^9$ \unitrho at the top of the crust from:
\begin{equation}
  \label{eq:coulomb}
  \Gamma \approx 173  \left( \frac{Z}{26} \right)^2  \left( \frac{56}{A} \right)^{1/3}  \left( \frac{3 \times 10^8 \text{K}}{T}\right)  \left( \frac{\rho}{10^9 \text{\unitrho} } \right)^{1/3},
\end{equation}
with $A$ the atomic mass of each ion and $\rho$ the density of matter, both enter by using the formula $\rho = n_i A m_u$.

Modes that exist in the ocean could penetrate into the crust since the boundary between these regions is in principle flexible \citep[][hereafter PB05a]{Piro05a}.
We solve a set of perturbation equations (described in the following section) upon a background which is cooling in the aftermath of a superburst. The background is treated as a series of static snapshots, which requires that the cooling timescale is much slower that the mode timescale.
Cooling follows the model of \citet{Cumming04a} and \citet{Keek15} described in Section~\ref{section:background}.
It is assumed that the amplitude of these perturbations decays to zero as the mode penetrates further into the crust.

Full solution of the modes of an accreting neutron star ocean requires: solving perturbations in spherical geometry, taking into account burning physics on the surface, accounting for magnetic field effects \citep[which could turn out to be dynamically important for some stars, especially if amplified during the burning process,][]{Cavecchi16}, allowing for the oblateness of the NS and its ocean due to rapid rotation, taking into account relativistic effects due to the strong gravitational field and rapid rotation, and coupling of all of the relevant layers from the crust to the photosphere (where radiative transport effects become important).

In what follows, we will neglect many of these effects and follow the approach of PB05b. This allows a direct comparison of our surface mode calculations for Carbon-triggered bursts with their surface mode calculations for H/He triggered bursts.

The general equation for gravitational acceleration inside a star, including the effects of general relativity, is:
\begin{equation}
  \label{eq:grav-gr}
  g(r) = \frac{Gm(r)}{r^2} \left( 1 - \frac{2 G m(r)}{r c^2} \right)^{-1/2} ,
\end{equation}
where $m(r)$ is the mass enclosed by a sphere of radius $r$, $G$ is the gravitational constant and $c$ is the speed of light. We assume a canonical NS of mass $M = 1.4$ M$_{\odot}$ and radius $R = 10$km, which results in surface gravity: $g \approx 2 \times 10^{14}$ \unitg.
Since the thickness of the ocean ($\Delta r \approx 10^4$ cm) is much less than the radius of a typical NSs we assume a constant surface gravity and that perturbations of the gravitational potential can be neglected (Cowling approximation). Relativistic effects due to frame dragging and redshift are discussed further in Section~\ref{section:results}. Though superbursts occur much deeper in a NS ocean than H/He bursts, the difference has little effect on the gravitation acceleration. This is also discussed further in Section~\ref{section:results}.

Mode solutions are found by perturbing the equations of continuity and momentum for a static background ($\bvec{v} = 0$). Rotational effects are assumed to be small in the unperturbed state, leaving only a radial dependence for the background.
Using $\delta$ for Eulerian perturbations, the perturbed versions of the continuity and momentum equations are written:
\begin{subequations}
  \begin{equation}
    \label{eq:continuity}
    \partial_t \delta \rho + \nabla \cdot \left( \rho \delta \bvec{v} \right) = 0,
  \end{equation}
  \begin{equation}
    \label{eq:euler}
    \parsub{t} \delta \bvec{v} =
    \frac{1}{\rho} \nabla \cdot \delta \boldsymbol{\sigma}
    - \frac{\delta \rho}{\rho} \bvec{g}
    - 2 \bvec{\Omega} \cross \delta \bvec{v},
  \end{equation}
\end{subequations}
with $\bvec{v}$ and $\rho$ the velocity and density of the fluid in a frame rotating with the star, and $\bvec{\Omega}$ and $\bvec{g}$ the rotation vector and gravitational field of the star. The stress tensor, $\boldsymbol{\sigma}$, is included at this stage and describes shearing relevant to modes that interact with the crust. Since there are no shear stresses in the ocean, the stress tensor simplifies to only having diagonal components equal to pressure: $\sigma_{ij} = - p \delta_{ij}$, making the term involving stress in Equation~\ref{eq:euler} simply the gradient of the pressure perturbation: $\nabla \cdot \delta \boldsymbol{\sigma} = - \nabla \delta p$.

We assume spins of less than $1$ kHz which is in line with observed burst oscillation frequencies and below breakup speed for the given NS. This meant that the centrifugal term could be neglected from Equation~\ref{eq:euler}. Modes are assumed to be adiabatic, which requires the thermal timescale to be much greater than the mode frequency: $t_{th} \gg 2 \pi / \omega$, which is the case in the deep ocean (at typical Carbon ignition depths, $t_{th} \sim 10^{5}$ s) and for the modes of interest here, which have $\omega / 2 \pi \sim 5 $ Hz. Using $\Delta$ for Lagrangian perturbations, the adiabatic approximation provides another equation for the density and pressure perturbations:
\begin{equation}
  \label{eq:adiabatic}
  \frac{\Delta p}{p} = \Gamma_1 \frac{\Delta \rho}{\rho},
\end{equation}
with $\Gamma_1 = \left( \partial \log p / \partial \log \rho \right)_s$, the adiabatic exponent.
Eulerian and Lagrangian perturbations of a quantity $Q$ are related using $\bvec{\xi}$, the Lagrangian fluid displacement, as:
\begin{equation}
  \label{eq:eulerlagrange}
  \Delta Q = \delta Q + \bvec{\xi} \cdot \nabla Q
\end{equation}
 Other useful quantities in this calculation are the pressure scale height:
\begin{equation}
  \label{eq:scaleheight}
  h = p / \rho g.
\end{equation}
and the \bruntvaisala frequency, $N$, which is given by:
\begin{equation}
  \label{eq:bruntvaisala}
  N^2 = - g \left( \partial_r \log \rho - \frac{1}{\Gamma_1} \partial_r \log p \right), 
\end{equation}
where $r$ is the radial coordinate in spherical geometry, and should be replaced by $z$ in plane parallel geometry.

In the next section we proceed to simplify these equations as far as possible, first deriving equations for modes that exist exclusively in the ocean in both spherical geometry and plane parallel geometry and then deriving equations for modes that penetrate into the crust only for the case of plane parallel geometry.

\subsection{Choosing a wave number}

A full derivation of the spherical perturbation equations in the ocean is given in \citet{Bildsten96} and \citet{Piro04}. Using spherical geometry, a perturbed quantity $Q$ is decomposed into solutions of the form: $\delta Q(r,\theta,\phi,t) = \delta Q(r, \theta) e^{im\phi - i\omega t}$ with $\omega$ the (angular) frequency of the wave, and $m$ the wave number. Perturbations of velocity are written in terms of the Lagrangian fluid displacement \citep{Friedman78}, which in the rotating frame and in the case of a static background is $\delta \bvec{v} = \difsub{t} \bvec{\xi}$.

Equations~\ref{eq:continuity}~and~\ref{eq:euler} are simplified by use of the \textit{Traditional Approximation} \citep{Eckart60}. Buoyancy and Coriolis forces in the radial component of Equation~\ref{eq:euler} are in competition, and their relative strengths can be estimated from slow rotating solutions for low frequency modes as: $\omega \ll N$, and $\xi_r / \xi_{\theta} \sim \Delta r / R$ where $\Delta r$ is the thickness of the layer. Dropping the Coriolis force term in the radial direction requires the spin frequency of the star to be $\Omega / 2 \pi \ll N^2 \Delta r / 2 \pi \omega R \sim 10^5$ Hz (using a 10 km NS, a $10^4$ cm thick layer, approximate frequency $\omega/2\pi = 10$ Hz, and $N / 2 \pi = 10^4$ Hz). This requirement is less restrictive than neglecting the centripetal force from Equation~\ref{eq:euler}. By assuming a low number of latitudinal nodes, terms of order $\xi_r / \xi_\phi$ in the $\phi$ component of Equation~\ref{eq:euler} may be dropped.

Given these assumptions, the ocean perturbation equations in spherical geometry are\footnote{Note that there is a misprint in \cite{Bildsten96} Equations (5-6), corrected in \cite{Maniopoulou04} Equations (55-56).}:
\begin{subequations}
  \begin{equation}
  \label{eq:perturbations-sph-a}
    \partial_r \left( \frac{\delta p}{p} \right) 
    = \frac{\xi_r}{gh} \left( \omega^2 - N^2 \right) + \frac{\delta p}{p} \frac{1}{h} \left( 1 - \frac{1}{\Gamma_1} \right) ,
  \end{equation}
  \begin{equation}
  \label{eq:perturbations-sph-b}
    \partial_r \xi_r + \frac{\xi_r}{h} \left( \frac{2 h}{R} - \frac{1}{\Gamma_1} \right) + \frac{ 1 }{\Gamma_1}\frac{ \delta p }{p}
    = - \frac{gh}{\omega^2 R^2} L_{\mu} \left( \frac{ \delta p }{p} \right),
  \end{equation}
\end{subequations}
where $L_\mu$ is a new operator which depends on the frequency and spin through the parameter $q = 2\Omega / \omega$, and wave number $m$. This operator is more conveniently written in terms of $\mu \equiv \cos \theta$ as:
\begin{multline}
  \label{eq:laplace-tidal}
  L_\mu \equiv
  \partial_\mu \left( \frac{1 - \mu^2}{1 - q^2 \mu^2} \partial_\mu \right)
  - \frac{m^2}{\left(1 - \mu^2\right) \left(1 - q^2 \mu^2\right)} \\
  - \frac{qm \left( 1 + q^2 \mu^2 \right)}{\left( 1 - q^2 \mu^2 \right)^{{2}}} ,
\end{multline}
and contains no radial dependence.
This operator appears in \textit{Laplace's Tidal Equation}; an eigenvalue problem $L_\mu \left( \delta p / p \right) = - \lambda \delta p / p$, the eigenfunctions of which are Hough functions. Methods of solution are outlined in \citet{Longuet-Higgins68}, \citet{Bildsten96} and \citet{Piro04}; we will not go into details here. Equations~\ref{eq:perturbations-sph-a}~and~\ref{eq:perturbations-sph-b} are separable, which means that solutions can be found by presupposing a mode and replacing the operator $L_\mu$ with the appropriate eigenvalue $-\lambda$, thus removing the angular dependence and leaving dependence on a single variable $r$. Solving the remaining pair of ordinary differential equations (hereafter, ODEs) requires knowledge of the background conditions. 

Rather than working in spherical geometry, we choose to solve the full set of equations including the crustal interaction in a plane parallel geometry, since this is what has been done in previous studies \citep[PB05a,b]{Bildsten95}, enabling an easy comparison of results.  The justification for making this switch of geometries is that the pressure scale height and thickness of the ocean, $h \sim 10^3$ cm, is much less than the radius of the star.

For a plane parallel geometry, a full derivation of the perturbation equations can be found in \citet{Bildsten95}. In this geometry there are only two variables: $z$ and $x$, where now $z$ is the vertical direction, and $x$ the horizontal displacement. Linking the calculations in plane parallel geometry to spherical geometry, $x$ follows lines of constant longitude and is related to small changes in $\theta$ in the spherical case by $\delta x \approx R \delta \theta$, while $z$ is mapped to $r$.
This means that a perturbed quantity $Q$ is decomposed as $\delta Q(z,x,t) = \delta Q(z) e^{ikx - i\omega t}$ with $\omega$ the angular frequency, and $k$ the wave number.

We start from Equations~\ref{eq:continuity}~and~\ref{eq:euler} and drop the Coriolis term. Dropping this term was performed in previous literature \citep{Bildsten95}, and will be necessary for crustal waves. It should limit spin frequencies to less than the frequency of the mode. The resulting equation, however, will be similar to those in spherical geometry. A case is made below that $k$ can be used to compensate for this simplification. The ocean perturbation equations in plane parallel geometry are:
\begin{subequations}
  \begin{equation}
    \label{eq:perturbations-ppl-a}
    \difsub{z} \left( \frac{\delta p}{p} \right) = \frac{\xi_z}{gh} \left( \omega^2 - N^2 \right) + \frac{\delta p}{p} \frac{1}{h} \left( 1 - \frac{1}{\Gamma_1} \right) ,
  \end{equation}
  \begin{equation}
    \label{eq:perturbations-ppl-b}
    \difsub{z} \xi_z - \frac{\xi_z}{\Gamma_1 h} + \frac{1}{\Gamma_1} \frac{\delta p}{p} = \frac{gh}{\omega^2} k^2 \frac{\delta p}{p} ,
  \end{equation}
\end{subequations}
which are ODEs in $z$. These equations are written in terms of the pressure perturbation, but using the $x$ component of the Equation~\ref{eq:euler} $\rho \omega^2 \xi_x = i k \delta p$ could be written in terms of $\xi_x$, the Lagrangian fluid displacement.

As can be seen, the two sets of equations for ocean perturbations, Equations~\ref{eq:perturbations-sph-a}~and~\ref{eq:perturbations-sph-b} and Equations~\ref{eq:perturbations-ppl-a}~and~\ref{eq:perturbations-ppl-b}, match when $k^2 = \lambda / R^2$ (neglecting a small term proportional to $\xi_r / R$). The spherical harmonic order and degree ($m$ and $l$) of the particular mode come through the constant $k$ or $\lambda$, which contain all the information involving the mode and, in the case of $\lambda$, rotation of the star. This means solving the perturbation equations in plane parallel geometry for some $k$ is equivalent to the spherical case for a suitable $\lambda = k^2 R^2$. It also means that when setting $k$ or $\lambda$ care must be taken to do so in a consistent manner.

\subsection{Crustal perturbations}
Crustal perturbation equations are derived in \cite{Bildsten95} and PB05a. They assume no shear stresses in the unperturbed background, as described above, and assume perturbations of the same form as the ocean case in plane parallel geometry ($\delta Q(z,x,t) = \delta Q(z) e^{ikx - i\omega t}$), with the wave vector choice governed by $k^2 = \lambda / R^2$. The equations are written here for completeness\footnote{These equations match \cite{Bildsten95} Equations (4.5a,b), and PB05a Equations (15-16).}:
\begin{subequations}
  \label{eq:crust-pert}
  \begin{multline}
    \difsub{z}^2 \xi_x 
    =
    \xi_x \left( \frac{\omega^2}{gh} - \frac{4 k^2 \mu}{3p} - \Gamma_1 k^2 \right)
    + \difsub{z}\xi_x \frac{1}{p} \difsub{z}\mu
    \\
    +
    ik \xi_z \left( \frac{1}{p} \difsub{z}\mu - \frac{1}{h} \right)
    + ik \difsub{z} \xi_z \left( \frac{\mu}{3p} + \Gamma_1 \right)
    ,
  \end{multline}
  \begin{multline}
    \difsub{z}^2 \xi_z \left( \Gamma_1 + \frac{4 \mu}{3 p} \right) 
    =
    ik \xi_x \left( \frac{1-\Gamma_1}{h} - \frac{2\difsub{z}\mu}{3p} + \difsub{z}\Gamma_1 \right)
    \\
    + ik \difsub{z} \xi_x \left( \frac{\mu}{3p} + \Gamma_1 \right)
    +
    \xi_z \left( \frac{\omega^2}{gh} - \frac{\mu k^2}{p} \right)
    \\
    + \difsub{z} \xi_z \left( \frac{4\difsub{z}\mu}{3p} - \frac{\Gamma_1}{h} + \difsub{z}\Gamma_1 \right)
    , 
  \end{multline}
\end{subequations}
where $\mu$ is the crystalline shear modulus in the crust and appears in the stress tensor $\boldsymbol{\sigma}$.
Note that these equations are written in terms of Lagrangian displacement in the $x$ direction, not pressure perturbations ($\xi_x$, not $\delta p$). \cite{Strohmayer91b} calculate $\mu$ as:
\begin{equation}
  \label{eq:shear_mod}
  \mu = \frac{0.1194}{1 + 0.595 ( 173/\Gamma )^2} \frac{ n_i (Ze)^2}{a} ,
\end{equation}
with $n_i$ the ion number density, $a$ the average ion spacing, and $\Gamma$ is the Coulomb coupling parameter. By assuming that the crust is that of a relativistic totally degenerate gas of electrons the appropriate equation of state relation can be used: $p = K n_i^{4/3}$, and the expression for shear modulus simplifies. This makes the shear modulus proportional to pressure, or $\mu / p$ approximately constant with depth, as:
\begin{equation}
  \label{eq:shear_mod_press}
  \frac{\mu}{p} = \frac{1.4 \times 10^{-2}}{1 + 0.595 ( 173/\Gamma )^2} \left( \frac{Z}{30} \right)^{2/3}.
\end{equation}
This approximation is reasonably accurate and simplifies the perturbation equations quite a bit, since derivatives of shear modulus can be related to derivatives of pressure.

\subsection{Motivating the transverse wave number}

The transverse wave number is chosen by considering Equation~\ref{eq:laplace-tidal}, which exhibits a variety of solutions including g-modes, Kelvin modes and r-modes \citep{Longuet-Higgins68}. \cite{Bildsten96} and \cite{Piro04} describe a dispersion relation between wave number, rotation, and frequency ($k$, $\Omega$, and $\omega$) for r-modes, which is used to set $k$ consistently with respect to rotation.
Only r-modes are considered here, however, since \cite{Heyl04} showed that buoyant r-modes with very low $m$ are the only solution to this equation that could be consistent with the basic properties of burst oscillations in H/He bursts.
PB05b adopted this restriction and focused their attention on  the $l=2$, $m=1$ r-mode in their calculation for H/He bursts. As we are comparing modes for superbursts to these results, we also choose this mode, for which $\lambda = 0.11$.

\subsection{Boundary conditions}

Boundary conditions for these modes are straightforward. At the outer boundary, in the shallow ocean, the Lagrangian pressure perturbations are set to zero ($\Delta p = 0$). At the inner boundary, deep within the crust ($\rho > 10^{11}$ \unitrho), the displacement perturbations are set to zero ($\xi_x , \xi_z = 0$).
Using these conditions, solutions are found using a shooting method. Solutions start in the ocean by solving Equations~\ref{eq:perturbations-ppl-a}~and~\ref{eq:perturbations-ppl-b} as far as the interface between ocean and crust.
Passing through the ocean-crust interface requires matching conditions, where the shear modulus, $\mu$, changes from zero to non-zero.
From here the crustal perturbation equations, Equations~\ref{eq:perturbations-ppl-a}~and~\ref{eq:perturbations-ppl-b}, are solved until the deep crust boundary depth.

There must be two constraints at the ocean-crust interface since the ocean perturbation equations form a pair of first order ODEs whereas the crust perturbation equations are a pair of second order ODEs.
Components of the stress tensor provide these constraints, namely: $\Delta \sigma_{zz}$ is assumed to be continuous across the interface, and $\Delta \sigma_{xz} = 0$ at the interface, resulting in conditions on the derivatives of $\xi_x, \xi_z$.
Across the interface it is assumed that the vertical component of the displacement perturbation, $\xi_z$, is continuous; however there is a discontinuity in transverse component, $\xi_x$, characterised by a parameter\footnote{Notation is changed from PB05a where the crustal discontinuity is $\lambda$, to avoid confusion with the eigenvalue of Laplace's Tidal Equation.}, $\chi$, where: $\xi_{x,\text{crust}} = \chi \xi_{x,\text{ocean}}$.

The only physical quantities under examination are the frequencies, $\omega$, which are not affected by the normalisation of the modes. A simple normalisation is, therefore, chosen: maximum perturbation is set to unity, $\mathrm{max} \left( \xi_x(\bvec{r}), \xi_z(\bvec{r}) \right) = 1$. Other possible choices might include, for example, constraints on the energy of the mode:
\begin{equation}
  \label{eq:mode-energy}
  E = 4 \pi R^2 \int \frac{1}{2} \rho \omega^2 \xi^2 ~ \mathrm{d} z
\end{equation}
as done in PB05b.

\section{Background Conditions} \label{section:background}
The evolving background is computed along similar lines to the model of \cite{Cumming04a} and \citet{Keek15}. The ocean is divided into two layers; a hot, shallow layer where heat from unstable nuclear burning has been deposited; and a cool, deeper layer that continues into, and is in thermal equilibrium with, the crust.
The boundary between the hot and the cold layers is the \textit{ignition depth}, which depends on the accretion rate, composition, and strength of \textit{shallow heating} in the outer layers of the crust (required to explain neutron star cooling curves and superburst recurrence times, see \citealt{Deibel16,Wijnands17}). The ignition depth dictates the cooling timescale of the burst, and for superbursts lies in the range of column depths\footnote{Defined in Section~\ref{section:cool-equns}}: $10^{11} - 10^{14}$ \unity \citep{Keek11}.
The bursting/hot layer cools with no extra sources of heat generation, and composition is fixed. Heat from this layer radiates from the star at the outer boundary and propagates into the cool ocean layer and crust throughout the course of the burst. 

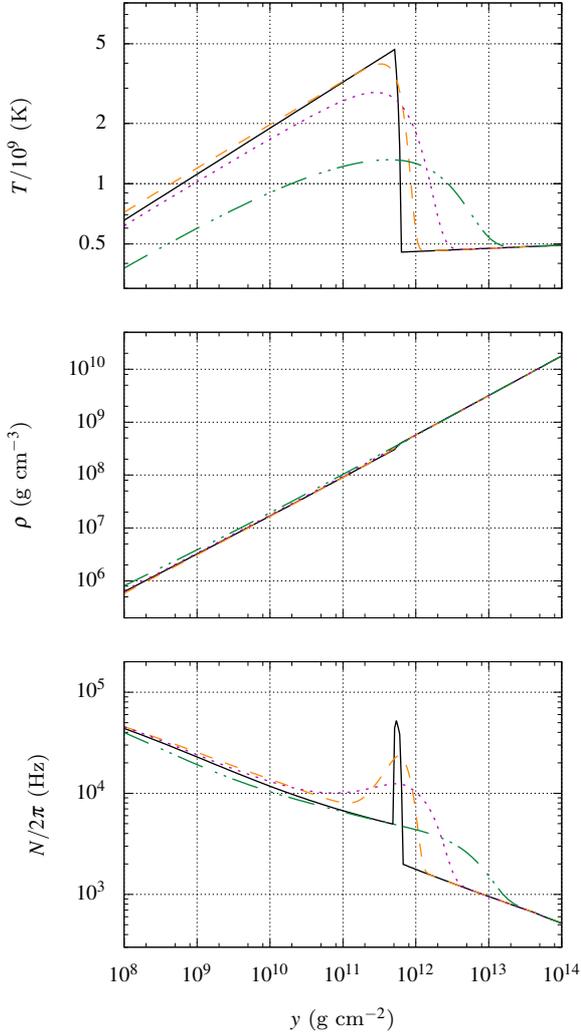
\begin{figure}
  \centering
  \input{images/cooling.tex}
  \caption{Cooling of the ocean in the aftermath of a burst using parameters: $\alpha = 0.23, \yign = 6 \times 10^{11}$ \unity, $\Tcrust = 5 \times 10^8$ K, and $\Tpeak = 4.8 \times 10^9$ K. Quantities plotted are temperature, density, and \bruntvaisala frequency. The profiles shown are for times $t = 0, 10^3, 10^4, 10^5$ secs in solid (black), dashed (orange), dotted (purple), and double-dot-dash (green) respectively. The full profiles range over depths $10^6 \unity$ to $10^{17}$ \unity. The crust starts at column depth $4.9 \times 10^{13}$ \unity.}
  \label{fig:cooling}
\end{figure}

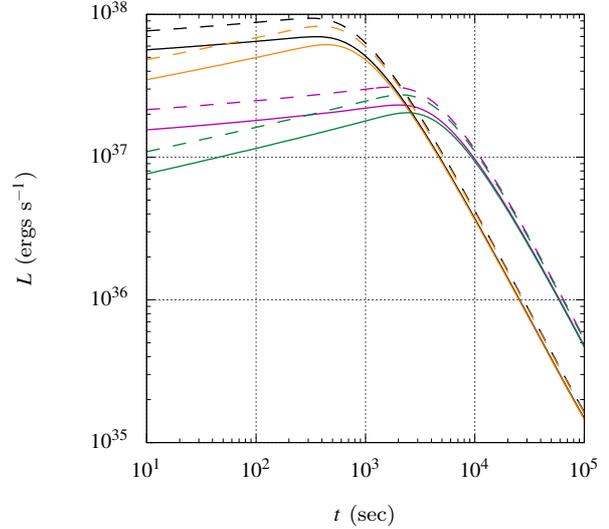
\begin{figure}
  \centering
  \input{images/luminosity.tex}
  \caption{Luminosity of several different parameter combinations for initial conditions as the layer cools. This figure can be compared to Figure 2 of \protect\cite{Keek15}, which compares the luminosity of this model to the observed luminosity during a superburst, removing the persistent and reflective components of the emission. Different slopes in the rising phase of the burst are for values of $\alpha$, steeper for greater $\alpha$ (in the colour version black and purple indicate $\alpha = 0.23 $, orange and green $\alpha = 0.25$). Different peak times correspond to different ignition depths, the earlier sets of peak for shallower depths $\yign = 2 \times 10^{11}$ \unity and later ones $6 \times 10^{11}$ \unity. Solid (dashed) lines are for $\Tpeak = \Tpeaka$ K ($\Tpeakb$ K). $\Tcrust$ is fixed at $\Tcrusta$ K as this parameter has little effect on luminosity.}
  \label{fig:luminosity}
\end{figure}

\subsection{Cooling equations}
\label{section:cool-equns}

Recall that rotational effects are neglected for the background so that quantities only have radial dependence. This means the Euler equation for the unperturbed quantities is simply hydrostatic equilibrium:
\begin{equation}
  \label{eq:hydrostatic}
  \partial_r p = - \rho g.
\end{equation}
Cooling in the wake of the  burst is calculated from the entropy equation:
\begin{equation}
  \label{eq:entropy}
  c_p \partial_t T = - \epsilon_\nu - \frac{1}{\rho} \partial_r F ,
\end{equation}
with $c_p$ the heat capacity at constant pressure, $\epsilon_{\nu}$ the neutrino energy loss rate, and $F$ the heat flux calculated according to the radiative transfer equation:
\begin{equation}
  \label{eq:rad-trans}
  F = - \frac{4ac T^3}{3\kappa \rho} \partial_r T ,
\end{equation}
with $a$ the radiation constant, $c$ the speed of light, and $\kappa$ the opacity. Cooling is calculated by assuming an initial temperature profile and allowing it to cool according to Equation~\ref{eq:entropy}.

Column depth, defined by $\mathrm{d} y = - \rho \mathrm{d} r$, is a more useful quantity to measure depth in the ocean: it represents the integrated surface mass density from the surface of the star to a point inside the star.
Since we have also assumed constant gravity, hydrostatic equilibrium reduces to: $p = g y$. Here we are using spherical coordinates, but the same equations are valid for plane parallel, substituting $r \rightarrow z$. Thus in all equations so far the spatial derivative is replaced by $y$ (and appropriate factors).

Solution of Equation~\ref{eq:entropy} uses the method of lines; spatial derivatives are calculated using finite difference methods upon a grid. This grid is chosen to be uniform in $\sinh^{-1} ( \log y / \ign{y} )$; this function concentrates points about the ignition depth $\ign{y}$. The resulting set of $T_i$ at points $y_i$ are variables in an ordinary differential equation in time which is solved using a stiff integrator. Gradients are calculated upon this grid using a fourth order finite difference scheme. For instance, the \bruntvaisala frequency, $N^2$, depends on the gradient of density which is calculated numerically. Boundary conditions are also required; the inner boundary condition holds temperature fixed deep within the crust, while the outer boundary holds the gradient of temperature fixed according to $\mathrm{d} \log T / \mathrm{d} \log p \propto 1/4$, which is the case for a radiatively diffusive atmosphere.

\subsection{Initial temperature profile}
\label{section:init-temp}

The model for the initial temperature profile is the same as in \cite{Keek15}. Heat in the shallow bursting layer radiates from the star at the outer boundary and propagates to the deep layer throughout the course of the burst. The cooling timescale is mostly dictated by $\ign{y}$.

The initial temperature profile of the shallower layer obeys a power law in column depth:
\begin{equation}
  \label{eq:init-temp}
  T = \Tpeak \left( \frac{y}{\yign} \right)^{\alpha} ,
\end{equation}
where $\Tpeak$ is the peak temperature of the burst (temperature at the ignition depth at the beginning of the burst), and $\alpha$ is the power law exponent. Choosing $\Tpeak$ as a parameter differs from \cite{Keek15}, where $\Tpeak$ is set by a normalisation condition, depending upon the energy released during the burst. This difference will be discussed further in Section~\ref{section:results}, where a comparison to observations is made.
The deeper layer which continues to the crust is at a near constant\footnote{The inner boundary acts as a heat bath where temperature remains fixed. For numerical reasons, there is a small, positive slope where the temperature changes by a factor of $1.1$ across the cool layer. Results are insensitive to this condition.} temperature $\Tcrust$.

These four parameters; $\alpha$, $\yign$, $\Tcrust$, and $\Tpeak$ are each varied over two possible values, listed in Table~\ref{table:init-cond}. This makes 16 different models for superbursts. An example of the temperature profile cooling in the aftermath of a superburst is given in Figure~\ref{fig:cooling}. As can be seen, it is relatively late into the burst ($\sim 10^3$ secs) that the temperature profile changes noticeably. The luminosity of the burst is also plotted in Figure~\ref{fig:luminosity}.

\begin{table}
  \centering
  \begin{tabular}{l | r r l}
    \hline
    Parameter & Value 1 & Value 2 & Unit \\
    \hline
    $\alpha$ & $\alphaa$ & $\alphab$ & -  \\
    $\yign$ & $\yigna$ & $\yignb$ & \unity  \\
    $\Tcrust$ & $\Tcrusta$ & $\Tcrustb$ & K  \\
    $\Tpeak$ & $\Tpeaka$ & $\Tpeakb$ & K  \\
    \hline
  \end{tabular}
  \caption{The different values each of the parameters can take in this study. For each parameter there is a choice of two values, so with four parameters this makes a total of 16 possible combinations, or models.}
  \label{table:init-cond}
\end{table}

\subsection{Composition and equation of state}
\label{section:comp-eos}

The most important compositional changes, $\sim$ 1s into the superburst, are electron captures on $^{56}$Ni, making $^{56}$Fe \citep[see fig. 6 of][for timescales]{Keek11}.
Since all the available nuclear energy is assumed to be used in the initial flash, the composition of the entire envelope is assumed to be the end product of the burning process: pure $^{56}$Fe \citep{Cumming04a}.
Throughout the burst the atmosphere is replaced with fresh H/He while the cooling is ongoing; this is not included in the calculation.

The equation of state (EOS) in the ocean and crust layers of a NS is that of a fully ionised non-ideal electron-ion plasma, as described in \citet{Potekhin10}.
This EOS is modeled using analytic approximations and fitting formulae for relevant thermodynamic processes, allowing efficient numerical implementation.
It is calculated with the aid of the freely available Ioffe Institute EOS code\footnote{\url{http://www.ioffe.ru/astro/EIP/eipintr.html} }.

Contributions from an ideal gas of electrons are included \citep{Chabrier98,Blinnikov96}, with non-ideal exchange-correlation contributions from the electron fluid \citep{Ichimaru87}.
The contributions of ideal ion-ion interactions are calculated in both the liquid \citep{Potekhin00} and crystal regimes \citep{Baiko01}.
Corrections for a strongly coupled Coulomb liquid \citep{DeWitt99} contain anharmonic corrections for classical and quantum regimes \citep{Baiko01,Farouki93,Carr61}, as well as corrections for different plasma regimes.
For mixtures of different ion species, the EOS is calculated using the linear mixing rule for the solid regime \citep{Potekhin09}.

The opacity is calculated from radiative and conductive components as: $1 / \kappa = 1 / \kappa_{\text{rad}} + 1 / \kappa_{\text{cond}}$. Radiative opacity is only dominant in the most shallow layer of the ocean where density is low, while conductive opacity dominates in the more dense layers. Since this calculation takes place in the deeper layers of the ocean, the main contribution to opacity is the conductive component.  Radiative opacity is calculated following \citet{Schatz99}, with contributions from electron scattering from \citet{Paczynski83b} and free-free absorption using the basic form of \citet{Clayton83}, {the Gaunt factor is calculated using fitting formulae devised by \citet{Schatz99} to match \citet{Itoh91}}; the total radiative opacity is calculated as the sum of these including a \textit{non-additivity} factor \citep{Potekhin01}.
Conductive opacity is calculated from fitting formulae in \citet{Potekhin99}, improving upon the basic form of \citet{Yakovlev80}. These formulae are valid in the ocean and crust of the star.
Neutrino cooling is important at very high temperatures, $T \geq 1$ GK, and are calculated using fitting formulae from \citet{Schinder87}.

\section{Results} \label{section:results}
In order to verify that the code written for this calculation worked as expected, we first reproduced results of previous calculations: the background evolution for both \citet{Cumming04a} and PB05b, and the mode calculations for PB05b. {Results are in good agreement, except for a $5$ second discrepancy in the time of transition from surface mode to crustal interface wave. This is due to the slower cooling rate for this calculation than PB05b, which acts to prolong drifts; and because the crustal wave calculated here is $\sim 1$ Hz lower than that of PB05b because of a slightly steeper gradient in the cool layer}. Having established the validity of this code, it was used to calculate the evolution of modes during a superburst.

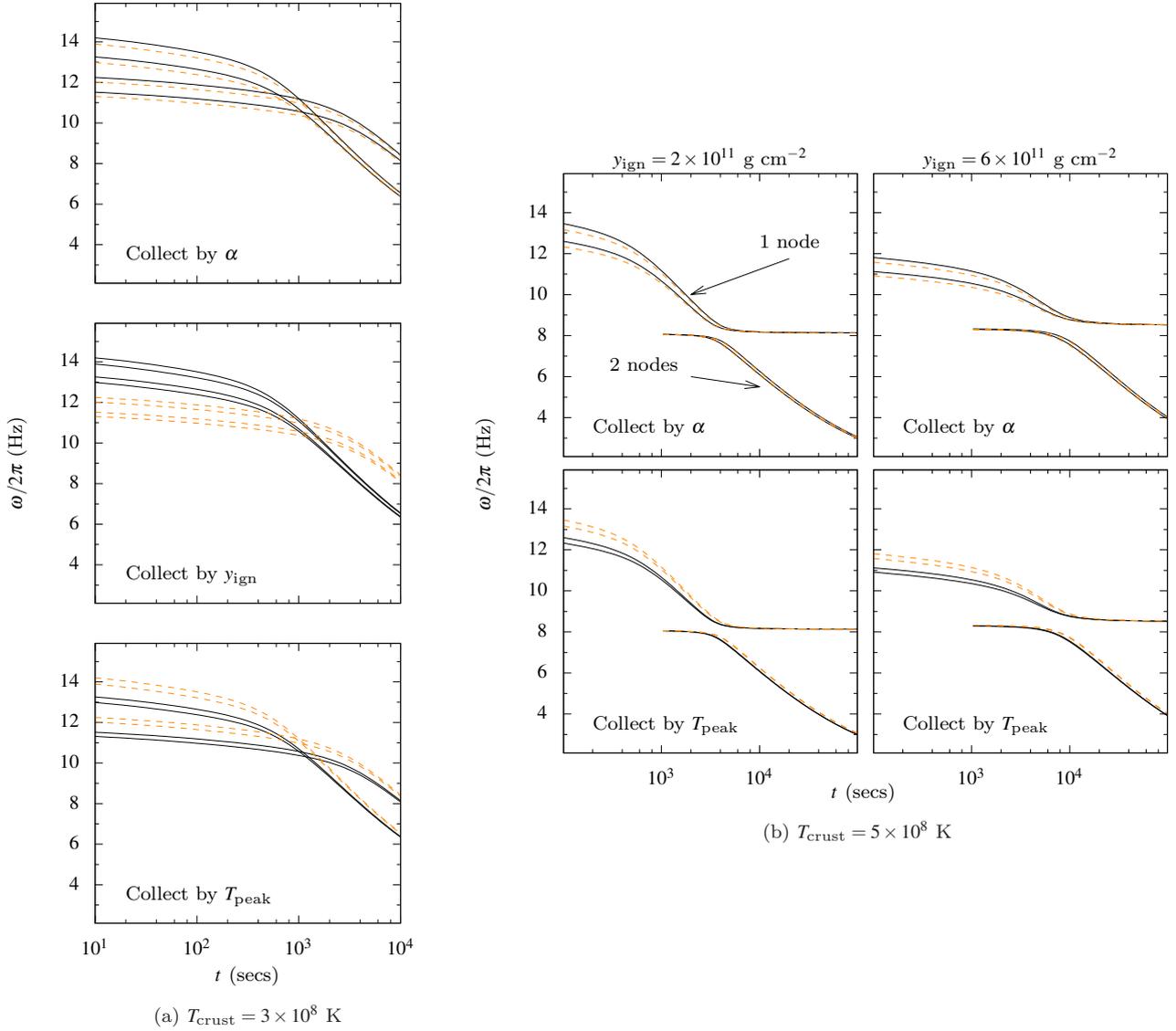
\begin{figure*}
  \centering
  \begin{minipage}{0.42\textwidth}
    \centering
    \input{images/eigenvalues-cool-crust.tex}
    \subcaption{$\Tcrust = \Tcrusta$ K}
    \label{fig:w-cool-crust}
  \end{minipage}
  \begin{minipage}{0.56\textwidth}
    \centering
    \input{images/eigenvalues-hot-crust.tex}
    \subcaption{$\Tcrust = \Tcrustb$ K}
    \label{fig:w-hot-crust}
  \end{minipage}
  \caption{Frequency evolution of r-mode and CIW solutions, in the rotating frame, for all models tested in this study. Figure~\ref{fig:w-cool-crust} shows solutions upon a background with a cool crust, while solutions in Figure~\ref{fig:w-hot-crust} are for a background with a hotter crust. Figure~\ref{fig:w-hot-crust} further separates background with different ignition depths, the left column containing shallower, and the right deeper, $\yign$.
    Over the time shown, cool crust solutions (Figure~\ref{fig:w-cool-crust}) do not undergo a transition to the CIW, whereas hot crust solutions (Figure~\ref{fig:w-hot-crust}) do demonstrate such a transition. For this reason a second set of lines are plotted in each panel of Figure~\ref{fig:w-hot-crust}; these are the CIWs for early times, and r-modes for late times. These extra solutions each have an extra node in their Lagrangian fluid displacement. For a discussion of whether or not this gives rise to an avoided crossing, see the text.
    Each panel shows solutions for several models, and collects models with the same parameter according to the line type (the common parameter is specified in the bottom left corner of the panel).
    For example, in the top panel of Figure~\ref{fig:w-cool-crust} solid-black lines are models with $\alpha = \alphaa$ and dashed-orange lines are models with $\alpha = \alphab$.
    Parameters belonging to the first column of Table~\ref{table:init-cond} (Value 1) are solid-black, and the second column dashed-orange.
    $\alpha$ and $\Tpeak$ have an effect on frequencies at early times, whilst $\yign$ affects frequency changes over the entire course of the burst. The strongest effect is due to $\Tcrust$, which dictates when the transition to the crustal interface wave occurs.
  }
  \label{fig:eigenvalues}
\end{figure*}

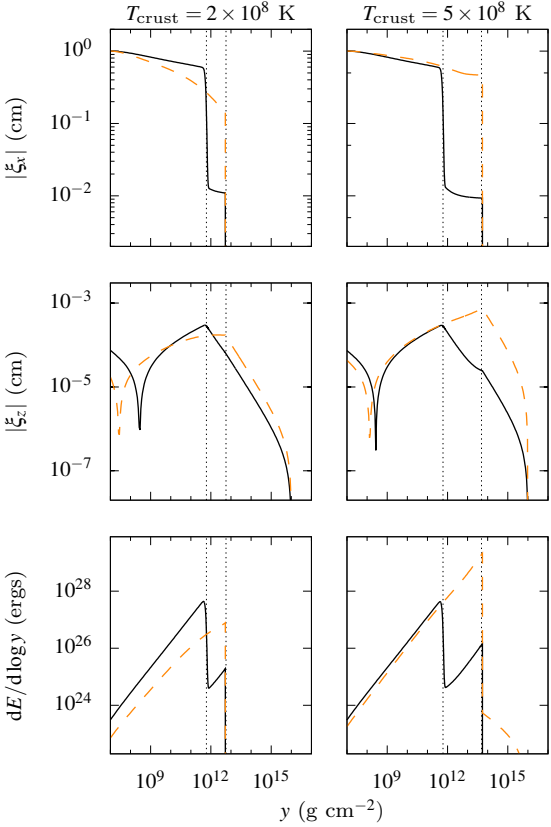
\begin{figure}
  \centering
  \input{images/eigenfunctions.tex}
  \caption{Structure of the mode solutions at different times for two background models with different values of $\Tcrust$. The left column shows solutions on a background with $\Tcrusta$ K (frequencies in Figure~\ref{fig:w-cool-crust}), and right column $\Tcrustb$ K (frequencies in Figure~\ref{fig:w-hot-crust}). Solid-black lines are for $10^2$ seconds after burst, and dashed-orange at $10^4$ seconds.
    The other parameters are the same: $\alpha = \alphaa$, $\ign{y} = \yignb$ \unity, and $\Tpeak = \Tpeaka$ K. Vertical lines on each panel represent the ignition depth and the ocean-crust transition depth.  The normalisation of these modes is not physical, hence the scale on the y-axis is arbitrary.
    At the later time for the hotter crust (right column, dashed-orange) the mode has undergone a transition to the CIW.
  }
  \label{fig:eigenfunctions}
\end{figure}

\subsection{Analytic frequency estimates}
\label{section:analytic-estimate}

Analytic estimates for the surface r-mode follow from the prescription for a shallow surface wave in a thin ocean limit given in \cite{Pedlosky87}:
\begin{equation}
  \label{eq:r-mode-estimate}
  \omega^2 \approx g \ign{h} k^2 \frac{\Delta \rho}{\rho}
\end{equation}
where $\ign{h}$ is the pressure scale height at the ignition site, and $k$ is the wavenumber.
As one moves across the ignition site, between the bursting and cool layers, there is a discontinuity in temperature and density, $\Delta T = T_b - T_c$ and $\Delta \rho = \rho_c - \rho_b$, while pressure remains the same ($\Delta p = 0$). The temperature of the bursting layer is greater than that of the cool layer, and vice versa for the density.

PB05b gave estimates for this frequency in the aftermath of a H/He burst by estimating pressure using an ideal gas of electrons. This approximation would be inappropriate for a superburst, where the pressure is due mainly to degenerate electrons. We approximate the pressure to be a sum of electron gases in the ultra-relativistic and ideal regimes:
\begin{equation}
  \label{eq:p-approx}
  p = K \left( \frac{\rho}{\mu_e} \right)^{4/3} + \frac{\rho}{\mu_e} \frac{k_B}{m_u} T.
\end{equation}
The degenerate component is the dominant pressure in both layers, however the ideal component of pressure is more significant in the hot bursting layer. A useful quantity to characterise which regime is dominant is the ratio of these two pressures:
\begin{equation}
  \label{eq:degen-fact}
  D_I \equiv \frac{p_{\text{ideal}}}{p_{\text{degen}}}
  = \frac{4k_B T}{E_F}
  \approx 0.15 \left( \frac{\rho / \mu_e}{ 10^8 \unitrho} \right)^{-1/3} \frac{T}{10^9 \text{K}}
\end{equation}
which depends weakly on density. We estimate this value on either side of the discontinuity, assuming that the density changes very little and remains at approximately $2 \times 10^8$ \unitrho. The temperature on the hot side is $\approx 4 \times 10^9$ K, which gives $D_I \approx 0.6$ and on the cool side the temperature is $\approx 4 \times 10^8$ K, so $D_I \approx 0.06$.

The relative change in density across the discontinuity can be obtained by equating the pressure on either side of the burst site, resulting in:
\begin{equation}
  \label{eq:den-disc}
  \frac{\Delta \rho}{\rho} \approx \frac{ \left( 1 + D_I \right) \Delta \mu / \mu + D_I \Delta T / T }{4 / 3  + D_I T_c / T_b}
\end{equation}
where $\mu_b$ and $\mu_c$ are mean molecular weight on the bursting and cool layers, $\Delta \mu / \mu \equiv \mu_c / \mu_b - 1$ is the fractional change in mean molecular weight across the discontinuity, and $D_I$ denotes the ratio of pressures in the bursting layer\footnote{There is a slight difference in the definition of $\Delta \rho / \rho$ to PB05b who use $\Delta \rho / \rho = 1 - \rho_b / \rho_c$, here we use $\Delta \rho / \rho = \rho_c / \rho_b - 1$}. {The importance of a composition change as compared to temperature change across the layer can be estimated from this result. Adding a quantity of carbon in the bursting layer so the mass fraction of carbon becomes $X_C = 0.2$, while keeping pure iron in the cool layer, means the ratio of the two terms in the numerator of Equation~\ref{eq:den-disc} is $3.6 \times 10^{-2}$; the first term is negligible in comparison to the second. In what follows, we proceed assuming the same composition in the bursting layer and cool layer, which makes
$\Delta \mu$ now zero, and the molecular weights equal, $\mu_b = \mu_c \equiv \mu_e$.}

The factor of $g \ign{h}$ can be calculated using the same equation of state, so that:
\begin{equation}
  \label{eq:hign-approx}
  g \ign{h} = \frac{\ign{p}}{\rho_b} = \frac{p_{\text{ideal}}}{\rho_b} \left( 1 + 1/D_I \right) = \frac{1}{\mu_e}\frac{k_{B}}{m_u} T_b \frac{1+D_I}{D_I}
\end{equation}
Finally, using $k^2 = \lambda / R^2$, as in Section~\ref{section:perturbations}, we find an estimate for frequency:
\begin{multline}
  \label{eq:freq-estimate}
  \frac{\omega}{2 \pi} \approx
  18 \text{Hz}
  \left( \frac{2Z}{A} \right)^{1/2}
  \left( \frac{T_b}{4 \times 10^{9} \text{K}} \right)^{1/2}
  \left( \frac{10 \text{km}}{R} \right)
  \\ \times
  \left( \frac{\lambda}{0.11} \right)^{1/2}
  \left( 1 - \frac{T_c}{T_b} \right)^{1/2}
  \left( \frac{1 + D_I}{1 + 3 D_I T_c / 4 T_b} \right)^{1/2} ,
\end{multline}
where $Z$ and $A$ are the average charge and mass of ions in the bursting layer. The final two factors in this expression are of order unity.

This result is similar to the expression for the frequency of r-modes during H/He bursts (PB05b, equation 3), the differences being a different temperature scale (PB05b use $10^9$ K) an extra factor for the significance of degeneracy, and terms in the final two factors for a different compositions in the bursting and cool layers which can be restored by using Equation~\ref{eq:den-disc}.
Taking the limit of only ideal gas contributions to the pressure, $D_I \rightarrow \infty$, results in the same expression as in PB05b.
This result predicts a strong dependence upon the temperature in the bursting layer, with increasing temperature resulting in increased frequency; and a slight dependence on the temperature of the cool layer. The temperature of the cool layer is only relevant when it is similar to that of the bursting layer, so as the bursting layer cools this term will become more important.
The parameter $\alpha$, the slope of the initial temperature profile, acts to control the temperature of the entire bursting layer. Since a decrease in $\alpha$ results in a higher average temperature for the layer (recall that the peak temperature is kept fixed), Equation~\ref{eq:freq-estimate} predicts an associated increase in frequency.

There is a dependence upon the ignition depth of a burst, through the final factor in the expression and implicitly through temperature. Given two bursts with different ignition depths (but the same temperatures in each of the respective layers) the deeper burst would have a smaller value for $D_I$, and thus a smaller frequency.

\subsection{Solutions}

Each model for the background described in Section~\ref{section:background} was used to solve Equation~\ref{eq:crust-pert}.
Rotating frame frequencies as a function of time after ignition are plotted in Figure~\ref{fig:eigenvalues}, with each panel of the figure demonstrating the effect of changing a single parameter across a series of different background models.
Example eigenfunctions of two models with differing values for $\Tcrust$ are shown in Figure~\ref{fig:eigenfunctions}, at $10^2$ and $10^4$ seconds after ignition.
The values for $\chi$, which measures the discontinuity in $\xi_x$ across the crust (see Section~\ref{section:perturbations}), are between $-10^{-4}$ and $- 10^{-5}$ for all models.

As found by PB05b, there are two families of solutions: a buoyant r-mode, concentrated in the surface layers; and a crustal interface wave (hereafter CIW). The frequency of the latter evolves very little as the superburst progresses.  The frequency of the former drops substantially as the envelope cools, and can (depending on the background parameters) cross the frequency of the CIW. The peak in energy of the buoyant r-mode eigenfunction also moves deeper as this occurs.
If the frequencies do cross, then before this occurs, solutions with a single node in $\xi_z$ and no nodes in $\xi_x$ are r-modes, while solutions with two nodes in $\xi_z$ and one in $\xi_x$ are CIWs. After the frequencies pass each other, the number of nodes switches between the two solutions.
For the duration of the calculation the buoyant r-mode solution on the background with a cooler crust remains an r-mode, while that on the background with a hotter crust has a frequency that crosses that of the CIW\footnote{For the  solution on the background with a cool crust, the transition occurs later in the burst than shown here (at $\sim 10^5$).}.
Whether or not the buoyant r-mode would actually transition to a CIW, in the cases where the frequencies cross, is an interesting question. \citet{Berkhout08} showed {that this transition cannot occur in the case of H/He triggered bursts, using the criterion that the cooling timescale must be greater than the inverse of the
  difference in frequency between the two branches at the avoided crossing, which is $\sim 1$ Hz (see Figure~\ref{fig:w-hot-crust}).
  This criterion is satisfied in the case of superbursts, the cooling timescale being of order $10^3$ seconds, so it is plausible that this transition may occur during superbursts}.

%% alpha
The parameter $\alpha$ only changes frequencies in the rising phase of the burst, with a larger value for $\alpha$ (a steeper slope in the initial temperature profile) resulting in a lower frequency, as predicted from the analytic estimate. The difference between different values of $\alpha$ is approximately $0.2$ Hz. At around the peak in luminosity the different temperature profiles converge and differences in frequency reduce. This is also the time at which frequencies resulting from different $\alpha$ start to converge, which is what one would expect from the analytic estimates, as the difference in temperature at the discontinuity decreases and reaches a similar value independent of $\alpha$.

%% y_ign
Shallower $\yign$ acts to increase frequencies substantially in the rising phase of the burst, by $2$ Hz, because a shallower $\yign$ results in a hotter bursting layer. This is captured in our model for initial temperature profile, but is also true for more extensive calculations of superbursts \citep[see fig. 3 of][]{Keek11}. Temperature stays higher for longer in deeper bursts (see the turnover time in Figure~\ref{fig:luminosity}) which is why the drop in frequencies occurs at later times for deeper bursts.

%% T_crust
The temperature of the crust has a small effect ($\sim 0.1$ Hz) on frequencies for early times, consistent with the analytic estimates. For late times this parameter still has little effect on frequencies ($\sim 0.3$ Hz).
Figure~\ref{fig:eigenfunctions} compares two models that only differ in $\Tcrust$ to demonstrate the transition from r-mode to CIW (should this transition occur). The reader should be reminded that these modes are normalised so that the maximum of $\xi_x^2 + \xi_z^2$ is one, so the value of the mode energy is arbitrary.

%% T_peak
Increasing $\Tpeak$ acts as to increase frequencies by $\sim 1$ Hz, which is as expected from the analytic estimates. As the layer cools, the difference in frequency between two models with different $\Tpeak$ diminishes. This is because the model with a higher $\Tpeak$ cools faster and results in a temperature profile similar to the smaller $\Tpeak$.

%% Damping times
Damping times are assumed to be long in comparison to the timescale of cooling, so are not considered here. Inspecting figures 6 and 8 of PB05b, which show the damping time and energy distribution of an r-mode during a H/He burst, it can be seen that the region where the damping time is comparable to cooling timescale occurs when the peak in energy of the mode is in the region $y < 10^{10} \unity$.  Since the peak in energy of the r-mode during a superburst is always deeper than this, we assume that damping time is similarly longer.

\subsection{Sensitivity to outer boundary and ocean-crust transition depth}

The depth of the outer boundary, $y_{\text{top}}$, was chosen at the location where the thermal time scale, $t_{\text{th}} = c_p y T / F$, is approximately equal to the timescale of the mode, $2 \pi / \omega$. This condition is due to the requirement that perturbations are adiabatic, which is true in the limit $\omega t_{\text{th}} / 2 \pi \gg 1$ \citep{Bildsten95}. For all models, and assuming a mode with frequency $\omega / 2 \pi \approx 10$, the value used in this calculation was $10^7$ \unity.

The sensitivity of frequencies to the depth of the outer boundary was tested. {At $t = 10^2$ and $10^4$ seconds after ignition, the frequency of the r-mode was measured for a range of different outer boundary depths: $10^6 - 10^8$ \unity. The frequency changes by $< 0.005$ Hz over the range of $y_{\text{top}}$, which is a small fraction of the average frequency ($<0.05\%$)}.
The frequency converges for shallower $y_{\text{top}}$; beyond $y_{\text{top}} \approx 10^7 \unity$ the adiabatic condition, $t_{\text{th}} > 2 \pi / \omega$, no longer holds, and residual perturbations this shallow do not affect the mode in the deeper ocean.

\begin{figure}
  \centering
  \input{images/ycrust-test.tex}
  \caption{Effect on frequencies of changing the depth of the ocean-crust transition. The two panels represent the calculations at $10^2$ and $10^4$ seconds. Frequencies are calculated over a range of $y_{\text{crust}}$ for two different crust temperatures $3 \times 10^{8}$ K in solid (black) and $5 \times 10^8$ K in dashed (orange). For early times the temperature of the crust and the depth are similarly important, however for late times the change in frequency due to the temperature of the crust is dwarfed by the effect of the depth. The ignition depth used in this calculation was $6 \times 10^{11}$ \unity.}
  \label{fig:ycrust-test}
\end{figure}
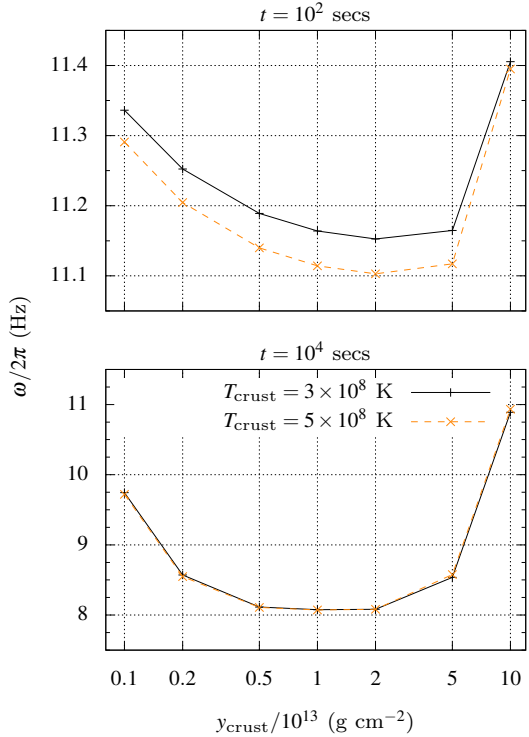

\begin{table}
  \centering
  \begin{tabular}{r r}
    \hline
    $\Tcrust$ (K) & $y_{\text{crust}}$ (\unity) \\
    \hline
    $3 \times 10^{8}$ & $5.6 \times 10^{12}$ \\
    $5 \times 10^{8}$ & $4.9 \times 10^{13}$ \\
    \hline
  \end{tabular}
  \caption{Depth of the ocean-crust interface for different crust temperatures.}
  \label{table:ycrust-test}
\end{table}

The depth of the ocean-crust transition, $y_{\text{crust}}$, is chosen at the point where $\Gamma = 173$, where $\Gamma$ is the Coulomb coupling parameter. This depth is dictated by the temperature in the crust where $y_{\text{crust}} = 5.6 \times 10^{12} \unity$ for $\Tcrust = 3 \times 10^8$ K, and $y_{\text{crust}} = 4.9 \times 10^{13} \unity$ for $\Tcrust = 5 \times 10^8$ K. This information is contained in Table~\ref{table:ycrust-test}. An interesting question is whether the frequency of the r-mode depends on $\Tcrust$ itself, or on $y_{\text{crust}}$ through this temperature.

Figure~\ref{fig:ycrust-test} shows frequencies for a range of different $y_{\text{crust}}$ between $10^{12} - 10^{14}$ \unity, at two times ($10^2, 10^4$ secs), for two different values of $\Tcrust$ ($3 \times 10^8$, $5 \times 10^8$ K), and for a set of the remaining background parameters ($\alpha, \yign$, and $\Tpeak$) where the buoyant r-mode frequency does not cross that of the CIW. For early times the frequency changes consistently by $\sim 0.05$ Hz between the two temperatures for a given depth, and by $\sim 0.15$ Hz for a given temperature over the range of different depths. It appears that the frequency depends on both $y_{\text{crust}}$ and $\Tcrust$, hence there is some intrinsic temperature dependence. This dependence, however, is very small since the frequency of the mode is dominated by shallower layers and the physics in the crust has a small effect. The same cannot be said for late times, where crustal interaction is more important. Frequencies change for different temperatures and the same $y_{\text{crust}}$ by $\sim 0.01$ Hz, while for the same temperature and different ocean-crust transition depths this change is $\sim 2$ Hz, or $\sim 3$ Hz at the extreme case.

\subsection{Comparison to 4U 1636-536}

A superburst from 4U 1636-536 was observed by \textit{RXTE} on 22 February 2001.
The observations lasted over the course of $\sim 10^4$ seconds, and exhibited oscillations over the 800 seconds after the burst peak \citep{Strohmayer02a}. The frequency of the oscillation remained very stable throughout cooling, with a drift of at most 0.1 Hz, and was a few Hz higher than the burst oscillation frequency seen in the H/He triggered bursts.
The drifts were consistent with those that would have been expected due to orbital Doppler shifts without requiring any intrinsic drift.

\cite{Keek15} were able to constrain some of the parameters of this superburst by using spectral analysis techniques and fitting parts of the light curve to a cooling model similar to the one used here. The parameters of the initial temperature profile were constrained to: initial slope $\alpha \approx 0.25$, ignition depth $\yign \approx 2 \times 10^{11}$ \unity, and energy deposited by the burst $E_{\text{nuc}} \approx 2.5 \times 10^{17}$ ergs g$^{-1}$. The parameter $E_{\text{nuc}}$ corresponds to $\Tpeak \approx 5 \times 10^9$ K for the given slope and ignition depth. 

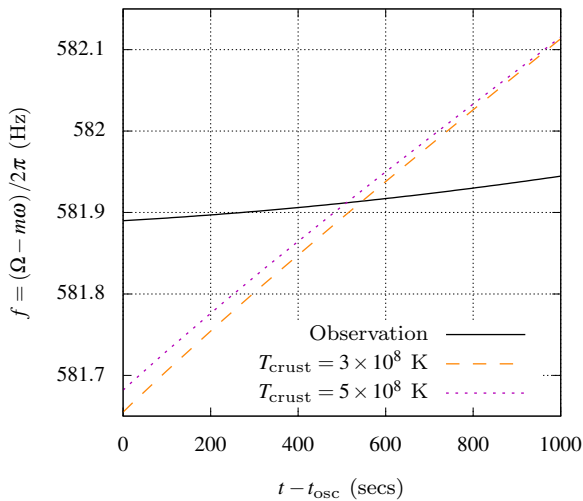
\begin{figure}
  \centering
  \input{images/4U-1636-536.tex}
  \caption{Observed superburst oscillation frequencies for 4U 1636-536 compared to frequencies of two models of buoyant $m=1$ r-modes (in the inertial frame, including orbital Doppler shifts). The fit to the data using an orbital Doppler shift model (drift by $<0.1$ Hz) from \protect\cite{Strohmayer02a} is solid (black) and the two models differ only in $\Tcrust$; dashed (orange) is $3 \times 10^8$ K, and dotted (purple) $5 \times 10^8$ K. The spin frequency is chosen to be $591.35$ Hz so as to best match the observed superburst oscillation frequency. The models assume superburst parameters, constrained by \protect\cite{Keek15}, as $\alpha = 0.25$, $\ign{y} = 2 \times 10^{11}$ \unity, and $\Tpeak = 5 \times 10^9$ K, however the temperature of the crust is unconstrained.}
  \label{fig:4U-1636-536}
\end{figure}

Figure~\ref{fig:4U-1636-536} compares the frequency evolution of the observations with that of buoyant r-modes in an inertial frame, including the effects of orbital Doppler shifts given in\citet{Strohmayer02a}, calculated using a background model that matches the constrained parameters.
The time covers the period over which oscillations are observed during the superburst; calculated modes are translated to match this time. From Figure~\ref{fig:luminosity}, the peak of the burst is at approximately $2.6 \times 10^2$ secs. 
Frequencies in an inertial frame are related to those in a co-rotating frame by: $\omega_{\text{i}} = \Omega - m\omega_{\text{rot}}$, neglecting relativistic effects (recall we calculated a mode with $m=1$). Two models are plotted, with different values of $\Tcrust$, since this is not constrained by \cite{Keek15}. This parameter would have more of an effect for late times in the superburst where the interaction with the crust is more important (especially if a transition to the CIW occurs), but by this time the oscillations in this particular observation we no longer detectable.
The spin frequency of the star is chosen to be $591.35$ Hz so as to best fit the observations.
This figure should be compared to figure 3 of \citet{Strohmayer02a}.

If the true spin frequency is 582 Hz, the asymptotic frequency of the H/He burst oscillations \citep{Strohmayer98a,Muno02b}, then the superburst oscillation model would give a frequency 10 Hz below the observed superburst oscillation frequency.
If the true spin frequency is 586 Hz, as suggested by the buoyant r-mode model of PB05b for the H/He burst oscillations, then the superburst oscillation model would predict a frequency 6 Hz below that observed.

\subsection{Constant gravity and GR effects on frequencies}
\label{section:freq-gr}

The assumption of using constant gravity is now considered. The calculation takes place over a wide range in column depth $10^7 - 10^{17}$ $\unity$ so it might be expected that the gravitational acceleration changes between the top and bottom of the layer. This assumption might be important when comparing modes between H/He bursts and superbursts, the ignition depths of which differ; $3 \times 10^{8}$ $\unity$  compared to  $10^{11} - 10^{12}$ \unity.

First we estimate the thickness and the mass of the layer in which the calculation takes place. These can be calculated using column depth with a crude estimate of the equation of state to be that of an ultra-relativistic electron gas, and that the radius is constant over the pressure range we consider:
\begin{subequations}
  \begin{equation}
    \label{eq:layerthickness}
    \Delta r 
    = \int^{y_{\text{base}}}_{y_{\text{top}}} \frac{\mathrm{d}y}{\rho} \approx 10^4 \text{cm} , 
  \end{equation}
  \begin{multline}
    \label{eq:layermass}
    \Delta m
    = \int^{y_{\text{base}}}_{y_{\text{top}}} 4 \pi r^2 \mathrm{d} y
    \approx \left( 6 \times 10^{-14} \msol \right) \\
    \times \left( \frac{R}{10 \text{km}} \right)^2 \frac{\Delta y}{10^{17} \text{g cm$^{-2}$ } } .
  \end{multline}
\end{subequations}
which means that difference in radius at the top or bottom of the layer is of order $ \Delta r / R \approx 0.01$, and the difference in enclosed mass is $\Delta m / M \approx 10^{-14}$.

Gravitational acceleration at a point $r$ inside the star is calculated from Equation~\ref{eq:grav-gr}. Using a constant $M$ in place of $m(r)$, and noting the Schwarzschild radius: $R_s = 2 G M / c^2$, the relative change in gravitational acceleration from the top of the layer to the base of the layer can be estimated by expanding the radius as $R \rightarrow R + \Delta r$ for the result:
\begin{equation}
  \label{eq:grav-error}
  \left| 1 - \frac{g(R+\Delta r)}{g(R)} \right| \approx
  \left( 2 + \frac{1}{2} \frac{R_s}{R} \right) \frac{\Delta r}{R} .
\end{equation}
Errors due to changes in gravity are of order 4\%. Analytic estimates of Equation~\ref{eq:r-mode-estimate} predict frequency scale with surface gravity as $\sqrt{g}$; errors in frequency due to changes in gravity should be of order 2 \%.

Using Equation~\ref{eq:layerthickness}, 
the difference in depth between the two ignition sites of H/He bursts and superbursts is estimated at $10^3$ cm. From Equation~\ref{eq:grav-error} the change in surface gravity is 0.4\%, also negligible. This enables us to compare results obtained for H/He bursts to superbursts.

\cite{Maniopoulou04} give general relativistic corrections to the traditional approximation due to frame-dragging and redshift, which depend on the compactness of the system and the modes under consideration. For the canonical NS mass and radius used here, frequency errors in the rotating frame of up to $15 \%$ are predicted for r-modes by including these effects, and $20 \%$ for Kelvin and g-modes. While these relativistic effects influence absolute frequencies of modes, the size of the drift during cooling is not affected (unless the mode undergoes a transition from one class of mode to another).

A $20 \%$ reduction in the frequency of buoyant r-modes, in the rotating frame, corresponds to a 2 Hz change. The spin frequency required to ensure a match to the observed superburst oscillation frequency would then be lower by 2 Hz. The problem with the frequency drifts, however, remains:  the model still predicts drifts of $0.3$ Hz over the duration of the observation, whereas the observed frequency drift is $\sim 0.04$ Hz, consistent with orbital Doppler shifts and no requirement for any intrinsic drift.

\section{Conclusion} \label{section:conclusion}
We have computed the frequencies of buoyant r-modes excited during superbursts. We found rotating frame frequencies to be greater, by $\sim 4$ Hz, than the frequencies of r-modes excited by H/He triggered Type I bursts as computed by PB05b, and drift by up to $\sim 10$ Hz in the aftermath of the burst.

According to the analytic approximations, Equation~\ref{eq:freq-estimate}, the frequency of the surface r-mode (in the rotating frame) scales with the temperature in the bursting layer. This explains the higher frequencies predicted for superbursts as compared to H/He burst since the temperatures are higher in superbursts.
Factors that might act to reduce the frequency are the degeneracy and composition. Limits on these, however, only act to reduce the frequency by $\sim 10$ \%. Should the same mode be present in both superbursts and H/He bursts, the frequency of the mode should be higher in the superbursts.

As can be seen from Figure~\ref{fig:4U-1636-536}, the frequencies of the mode calculated here do not match those of 4U 1636-536.
A spin frequency of $591.35$ Hz would be required for the mode model to match the superburst oscillation observations, whereas using a mode model for H/He triggered burst oscillations of the same system would predict a spin frequency for the NS not greater than $586$ Hz.
This is a large discrepancy, should burst oscillations and superburst oscillations both be explained by a mode model.  The frequency drifts predicted by the superburst oscillation model are also larger than those observed. The observations can be explained using Doppler shifts without any intrinsic frequency drift, whereas our model predicts an observed drift $\sim$ 1.5 Hz during the time period in which the oscillations were observed.

Another possible explanation of these oscillations is that the frequency comes directly from a CIW, the frequency of which would be stable throughout the burst.  An explanation of how this wave could couple to observable variations in the photosphere is however problematic, given how small the energy perturbations are in that region in comparison to the crust.  A more accurate cooling model that accounts of the physics of the crust (equation of state, free neutrons, changes in composition), and more comprehensive solution of the perturbation equations in spherical geometry would be required, which is beyond the scope of this paper.
The possibility of a transition from r-mode to CIW at later times certainly needs to be studied in greater detail. The avoided crossing and associated change of mode type, ruled out for H/He bursts by \citet{Berkhout08}, cannot be ruled out for superbursts since the cooling timescale is sufficiently long, and the frequency difference between the two branches sufficiently small. The original analysis also had caveats, whose applicability remains to be explored \citep[section 4.1]{Berkhout08}.

There is physics missing from this mode calculation that could have significant effects on the frequencies of modes. This includes the effects of magnetic fields, ongoing nuclear burning throughout the course of the burst (this would strongly affect temperatures which have the largest effect on frequencies), and general relativistic effects (although estimates of these are given in Section~\ref{section:freq-gr} and are estimated to be of the order 2 \%).  The influence of the photosphere and outer layers of the atmosphere on frequencies is also not taken into account. Given the observed emission emerges from this region of the star, any model proposed to explain oscillations on the surface of the star must eventually consider how the mechanism would be affected by the photosphere.

The surface modes found here are very sensitive to the background model and therefore the underlying physics of superbursts. Different initial temperature models can change frequencies by up to $\sim 4$ Hz (see Figure~\ref{fig:eigenvalues}). This high sensitivity means that if modes do exist during superbursts and these bursts can be observed in high time resolution they will be an excellent probe of the physics of the deep ocean. The next generation of large high time resolution X-ray satellites such as the proposed eXTP \citep{eXTP} and Strobe-X \citep{Strobe-X}, may be able to detect these differences.

\subsection*{Acknowledgments}

FRNC, ALW and FG acknowledge support from ERC Starting Grant No. 639217 CSINEUTRONSTAR (PI Watts). YC is supported by the European Union Horizon 2020 research and innovation programme under the Marie Sklodowska-Curie Global Fellowship grant agreement No 703916.  This paper benefited from NASA's Astrophysics Data System. We thank the Ioffe Institute for making their equation of state routines publicly available. This collaboration was enabled by the National Science Foundation under Grant No. PHY-1430152 (JINA Center for the Evolution of the Elements).  We would like to thank Hendrik Schatz and Alexander Heger for useful discussions.

\bibliographystyle{mnras}
\bibliography{sbosc}

%%%%%%%%%%%%%%%%%%%%%%%%%%%%%%%%%%%%%%%%%%%%%%%%%%
%%%%%%%%%%%%%%%%%%%% APPENDIX %%%%%%%%%%%%%%%%%%%%
% \appendix

% Don't change these lines
\bsp	% typesetting comment
\label{lastpage}
\end{document}

%% file: images/cooling.tex
% GNUPLOT: LaTeX picture with Postscript
\begingroup
\newcommand{\tic}[0]{\tiny} \newcommand{\ax}[0]{\normalsize}
  \makeatletter
  \providecommand\color[2][]{%
    \GenericError{(gnuplot) \space\space\space\@spaces}{%
      Package color not loaded in conjunction with
      terminal option `colourtext'%
    }{See the gnuplot documentation for explanation.%
    }{Either use 'blacktext' in gnuplot or load the package
      color.sty in LaTeX.}%
    \renewcommand\color[2][]{}%
  }%
  \providecommand\includegraphics[2][]{%
    \GenericError{(gnuplot) \space\space\space\@spaces}{%
      Package graphicx or graphics not loaded%
    }{See the gnuplot documentation for explanation.%
    }{The gnuplot epslatex terminal needs graphicx.sty or graphics.sty.}%
    \renewcommand\includegraphics[2][]{}%
  }%
  \providecommand\rotatebox[2]{#2}%
  \@ifundefined{ifGPcolor}{%
    \newif\ifGPcolor
    \GPcolortrue
  }{}%
  \@ifundefined{ifGPblacktext}{%
    \newif\ifGPblacktext
    \GPblacktextfalse
  }{}%
  % define a \g@addto@macro without @ in the name:
  \let\gplgaddtomacro\g@addto@macro
  % define empty templates for all commands taking text:
  \gdef\gplbacktext{}%
  \gdef\gplfronttext{}%
  \makeatother
  \ifGPblacktext
    % no textcolor at all
    \def\colorrgb#1{}%
    \def\colorgray#1{}%
  \else
    % gray or color?
    \ifGPcolor
      \def\colorrgb#1{\color[rgb]{#1}}%
      \def\colorgray#1{\color[gray]{#1}}%
      \expandafter\def\csname LTw\endcsname{\color{white}}%
      \expandafter\def\csname LTb\endcsname{\color{black}}%
      \expandafter\def\csname LTa\endcsname{\color{black}}%
      \expandafter\def\csname LT0\endcsname{\color[rgb]{1,0,0}}%
      \expandafter\def\csname LT1\endcsname{\color[rgb]{0,1,0}}%
      \expandafter\def\csname LT2\endcsname{\color[rgb]{0,0,1}}%
      \expandafter\def\csname LT3\endcsname{\color[rgb]{1,0,1}}%
      \expandafter\def\csname LT4\endcsname{\color[rgb]{0,1,1}}%
      \expandafter\def\csname LT5\endcsname{\color[rgb]{1,1,0}}%
      \expandafter\def\csname LT6\endcsname{\color[rgb]{0,0,0}}%
      \expandafter\def\csname LT7\endcsname{\color[rgb]{1,0.3,0}}%
      \expandafter\def\csname LT8\endcsname{\color[rgb]{0.5,0.5,0.5}}%
    \else
      % gray
      \def\colorrgb#1{\color{black}}%
      \def\colorgray#1{\color[gray]{#1}}%
      \expandafter\def\csname LTw\endcsname{\color{white}}%
      \expandafter\def\csname LTb\endcsname{\color{black}}%
      \expandafter\def\csname LTa\endcsname{\color{black}}%
      \expandafter\def\csname LT0\endcsname{\color{black}}%
      \expandafter\def\csname LT1\endcsname{\color{black}}%
      \expandafter\def\csname LT2\endcsname{\color{black}}%
      \expandafter\def\csname LT3\endcsname{\color{black}}%
      \expandafter\def\csname LT4\endcsname{\color{black}}%
      \expandafter\def\csname LT5\endcsname{\color{black}}%
      \expandafter\def\csname LT6\endcsname{\color{black}}%
      \expandafter\def\csname LT7\endcsname{\color{black}}%
      \expandafter\def\csname LT8\endcsname{\color{black}}%
    \fi
  \fi
    \setlength{\unitlength}{0.0500bp}%
    \ifx\gptboxheight\undefined%
      \newlength{\gptboxheight}%
      \newlength{\gptboxwidth}%
      \newsavebox{\gptboxtext}%
    \fi%
    \setlength{\fboxrule}{0.5pt}%
    \setlength{\fboxsep}{1pt}%
\begin{picture}(4320.00,8640.00)%
      \csname LTb\endcsname%
      \put(2160,8420){\makebox(0,0){\strut{}}}%
    \gplgaddtomacro\gplbacktext{%
      \csname LTb\endcsname%
      \put(660,5947){\makebox(0,0)[r]{\strut{}$0.5$}}%
      \csname LTb\endcsname%
      \put(660,6400){\makebox(0,0)[r]{\strut{}$1$}}%
      \csname LTb\endcsname%
      \put(660,6853){\makebox(0,0)[r]{\strut{}$2$}}%
      \csname LTb\endcsname%
      \put(660,7452){\makebox(0,0)[r]{\strut{}$5$}}%
      \csname LTb\endcsname%
      \put(792,5393){\makebox(0,0){\strut{}}}%
      \csname LTb\endcsname%
      \put(1336,5393){\makebox(0,0){\strut{}}}%
      \csname LTb\endcsname%
      \put(1880,5393){\makebox(0,0){\strut{}}}%
      \csname LTb\endcsname%
      \put(2424,5393){\makebox(0,0){\strut{}}}%
      \csname LTb\endcsname%
      \put(2967,5393){\makebox(0,0){\strut{}}}%
      \csname LTb\endcsname%
      \put(3511,5393){\makebox(0,0){\strut{}}}%
      \csname LTb\endcsname%
      \put(4055,5393){\makebox(0,0){\strut{}}}%
    }%
    \gplgaddtomacro\gplfronttext{%
      \csname LTb\endcsname%
      \put(22,6686){\rotatebox{-270}{\makebox(0,0){\strut{}$T / 10^9$ (K)}}}%
      \put(2423,5327){\makebox(0,0){\strut{}}}%
      \put(2423,7649){\makebox(0,0){\strut{}}}%
    }%
    \gplgaddtomacro\gplbacktext{%
      \csname LTb\endcsname%
      \put(660,3414){\makebox(0,0)[r]{\strut{}$10^{6}$}}%
      \csname LTb\endcsname%
      \put(660,3812){\makebox(0,0)[r]{\strut{}$10^{7}$}}%
      \csname LTb\endcsname%
      \put(660,4210){\makebox(0,0)[r]{\strut{}$10^{8}$}}%
      \csname LTb\endcsname%
      \put(660,4607){\makebox(0,0)[r]{\strut{}$10^{9}$}}%
      \csname LTb\endcsname%
      \put(660,5005){\makebox(0,0)[r]{\strut{}$10^{10}$}}%
      \csname LTb\endcsname%
      \put(792,2916){\makebox(0,0){\strut{}}}%
      \csname LTb\endcsname%
      \put(1336,2916){\makebox(0,0){\strut{}}}%
      \csname LTb\endcsname%
      \put(1880,2916){\makebox(0,0){\strut{}}}%
      \csname LTb\endcsname%
      \put(2424,2916){\makebox(0,0){\strut{}}}%
      \csname LTb\endcsname%
      \put(2967,2916){\makebox(0,0){\strut{}}}%
      \csname LTb\endcsname%
      \put(3511,2916){\makebox(0,0){\strut{}}}%
      \csname LTb\endcsname%
      \put(4055,2916){\makebox(0,0){\strut{}}}%
    }%
    \gplgaddtomacro\gplfronttext{%
      \csname LTb\endcsname%
      \put(22,4209){\rotatebox{-270}{\makebox(0,0){\strut{}$\rho$ (g cm$^{-3}$)}}}%
      \put(2423,2850){\makebox(0,0){\strut{}}}%
      \put(2423,5173){\makebox(0,0){\strut{}}}%
    }%
    \gplgaddtomacro\gplbacktext{%
      \csname LTb\endcsname%
      \put(660,1057){\makebox(0,0)[r]{\strut{}$10^{3}$}}%
      \csname LTb\endcsname%
      \put(660,1817){\makebox(0,0)[r]{\strut{}$10^{4}$}}%
      \csname LTb\endcsname%
      \put(660,2577){\makebox(0,0)[r]{\strut{}$10^{5}$}}%
      \csname LTb\endcsname%
      \put(792,440){\makebox(0,0){\strut{}$10^{8}$}}%
      \csname LTb\endcsname%
      \put(1336,440){\makebox(0,0){\strut{}$10^{9}$}}%
      \csname LTb\endcsname%
      \put(1880,440){\makebox(0,0){\strut{}$10^{10}$}}%
      \csname LTb\endcsname%
      \put(2424,440){\makebox(0,0){\strut{}$10^{11}$}}%
      \csname LTb\endcsname%
      \put(2967,440){\makebox(0,0){\strut{}$10^{12}$}}%
      \csname LTb\endcsname%
      \put(3511,440){\makebox(0,0){\strut{}$10^{13}$}}%
      \csname LTb\endcsname%
      \put(4055,440){\makebox(0,0){\strut{}$10^{14}$}}%
    }%
    \gplgaddtomacro\gplfronttext{%
      \csname LTb\endcsname%
      \put(154,1733){\rotatebox{-270}{\makebox(0,0){\strut{}$N / 2 \pi$ (Hz)}}}%
      \put(2423,110){\makebox(0,0){\strut{}$y$ (g cm$^{-2}$)}}%
      \put(2423,2696){\makebox(0,0){\strut{}}}%
    }%
    \gplbacktext
    \put(0,0){\includegraphics{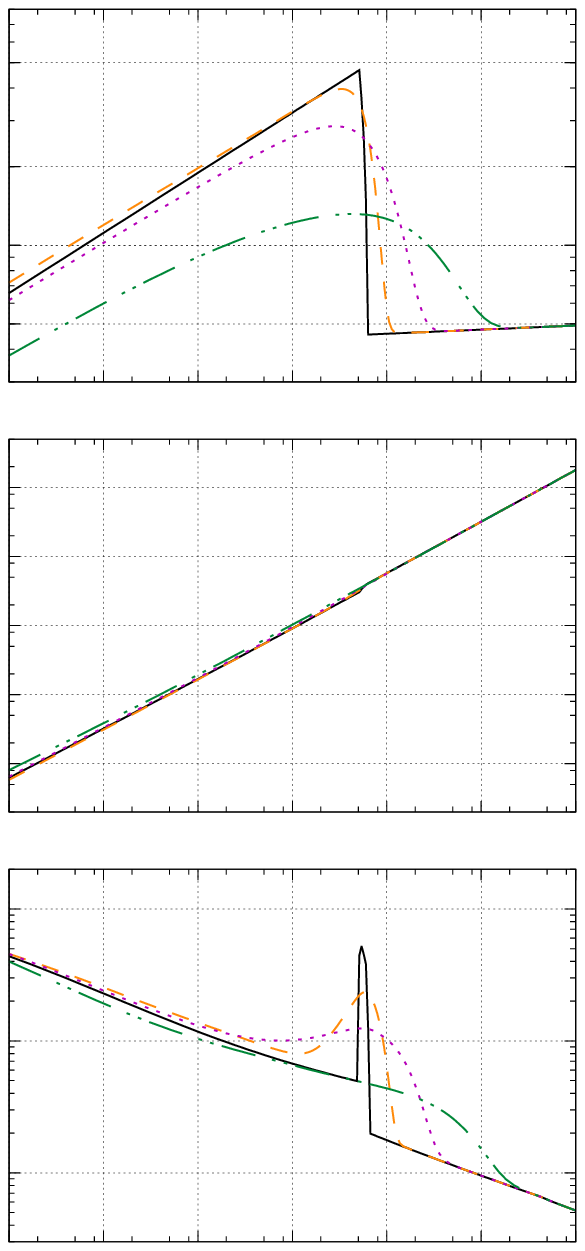}}%
    \gplfronttext
  \end{picture}%
\endgroup

%% file: images/luminosity.tex
% GNUPLOT: LaTeX picture with Postscript
\begingroup
\newcommand{\tic}[0]{\tiny} \newcommand{\ax}[0]{\normalsize}
  \makeatletter
  \providecommand\color[2][]{%
    \GenericError{(gnuplot) \space\space\space\@spaces}{%
      Package color not loaded in conjunction with
      terminal option `colourtext'%
    }{See the gnuplot documentation for explanation.%
    }{Either use 'blacktext' in gnuplot or load the package
      color.sty in LaTeX.}%
    \renewcommand\color[2][]{}%
  }%
  \providecommand\includegraphics[2][]{%
    \GenericError{(gnuplot) \space\space\space\@spaces}{%
      Package graphicx or graphics not loaded%
    }{See the gnuplot documentation for explanation.%
    }{The gnuplot epslatex terminal needs graphicx.sty or graphics.sty.}%
    \renewcommand\includegraphics[2][]{}%
  }%
  \providecommand\rotatebox[2]{#2}%
  \@ifundefined{ifGPcolor}{%
    \newif\ifGPcolor
    \GPcolortrue
  }{}%
  \@ifundefined{ifGPblacktext}{%
    \newif\ifGPblacktext
    \GPblacktextfalse
  }{}%
  % define a \g@addto@macro without @ in the name:
  \let\gplgaddtomacro\g@addto@macro
  % define empty templates for all commands taking text:
  \gdef\gplbacktext{}%
  \gdef\gplfronttext{}%
  \makeatother
  \ifGPblacktext
    % no textcolor at all
    \def\colorrgb#1{}%
    \def\colorgray#1{}%
  \else
    % gray or color?
    \ifGPcolor
      \def\colorrgb#1{\color[rgb]{#1}}%
      \def\colorgray#1{\color[gray]{#1}}%
      \expandafter\def\csname LTw\endcsname{\color{white}}%
      \expandafter\def\csname LTb\endcsname{\color{black}}%
      \expandafter\def\csname LTa\endcsname{\color{black}}%
      \expandafter\def\csname LT0\endcsname{\color[rgb]{1,0,0}}%
      \expandafter\def\csname LT1\endcsname{\color[rgb]{0,1,0}}%
      \expandafter\def\csname LT2\endcsname{\color[rgb]{0,0,1}}%
      \expandafter\def\csname LT3\endcsname{\color[rgb]{1,0,1}}%
      \expandafter\def\csname LT4\endcsname{\color[rgb]{0,1,1}}%
      \expandafter\def\csname LT5\endcsname{\color[rgb]{1,1,0}}%
      \expandafter\def\csname LT6\endcsname{\color[rgb]{0,0,0}}%
      \expandafter\def\csname LT7\endcsname{\color[rgb]{1,0.3,0}}%
      \expandafter\def\csname LT8\endcsname{\color[rgb]{0.5,0.5,0.5}}%
    \else
      % gray
      \def\colorrgb#1{\color{black}}%
      \def\colorgray#1{\color[gray]{#1}}%
      \expandafter\def\csname LTw\endcsname{\color{white}}%
      \expandafter\def\csname LTb\endcsname{\color{black}}%
      \expandafter\def\csname LTa\endcsname{\color{black}}%
      \expandafter\def\csname LT0\endcsname{\color{black}}%
      \expandafter\def\csname LT1\endcsname{\color{black}}%
      \expandafter\def\csname LT2\endcsname{\color{black}}%
      \expandafter\def\csname LT3\endcsname{\color{black}}%
      \expandafter\def\csname LT4\endcsname{\color{black}}%
      \expandafter\def\csname LT5\endcsname{\color{black}}%
      \expandafter\def\csname LT6\endcsname{\color{black}}%
      \expandafter\def\csname LT7\endcsname{\color{black}}%
      \expandafter\def\csname LT8\endcsname{\color{black}}%
    \fi
  \fi
    \setlength{\unitlength}{0.0500bp}%
    \ifx\gptboxheight\undefined%
      \newlength{\gptboxheight}%
      \newlength{\gptboxwidth}%
      \newsavebox{\gptboxtext}%
    \fi%
    \setlength{\fboxrule}{0.5pt}%
    \setlength{\fboxsep}{1pt}%
\begin{picture}(4320.00,4320.00)%
    \gplgaddtomacro\gplbacktext{%
      \csname LTb\endcsname%
      \put(660,660){\makebox(0,0)[r]{\strut{}$10^{35}$}}%
      \csname LTb\endcsname%
      \put(660,1733){\makebox(0,0)[r]{\strut{}$10^{36}$}}%
      \csname LTb\endcsname%
      \put(660,2806){\makebox(0,0)[r]{\strut{}$10^{37}$}}%
      \csname LTb\endcsname%
      \put(660,3879){\makebox(0,0)[r]{\strut{}$10^{38}$}}%
      \csname LTb\endcsname%
      \put(792,440){\makebox(0,0){\strut{}$10^{1}$}}%
      \csname LTb\endcsname%
      \put(1608,440){\makebox(0,0){\strut{}$10^{2}$}}%
      \csname LTb\endcsname%
      \put(2424,440){\makebox(0,0){\strut{}$10^{3}$}}%
      \csname LTb\endcsname%
      \put(3239,440){\makebox(0,0){\strut{}$10^{4}$}}%
      \csname LTb\endcsname%
      \put(4055,440){\makebox(0,0){\strut{}$10^{5}$}}%
    }%
    \gplgaddtomacro\gplfronttext{%
      \csname LTb\endcsname%
      \put(-110,2269){\rotatebox{-270}{\makebox(0,0){\strut{}$L$ (ergs s$^{-1}$)}}}%
      \put(2423,110){\makebox(0,0){\strut{}$t$ (sec)}}%
    }%
    \gplbacktext
    \put(0,0){\includegraphics{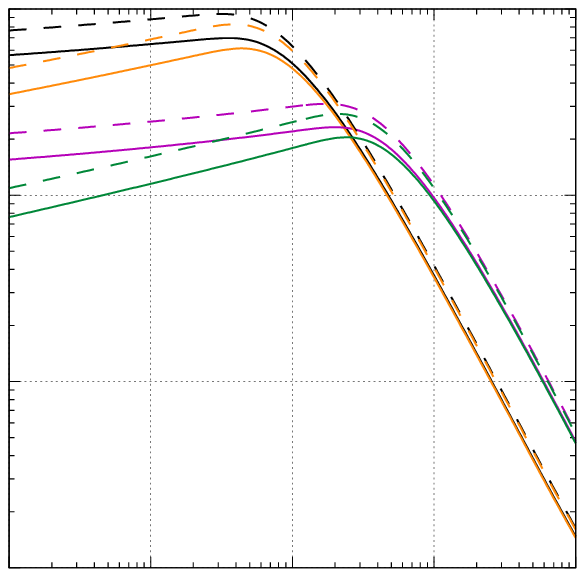}}%
    \gplfronttext
  \end{picture}%
\endgroup

%% file: images/eigenvalues-cool-crust.tex
% GNUPLOT: LaTeX picture with Postscript
\begingroup
\newcommand{\tic}[0]{\tiny} \newcommand{\ax}[0]{\normalsize}
  \makeatletter
  \providecommand\color[2][]{%
    \GenericError{(gnuplot) \space\space\space\@spaces}{%
      Package color not loaded in conjunction with
      terminal option `colourtext'%
    }{See the gnuplot documentation for explanation.%
    }{Either use 'blacktext' in gnuplot or load the package
      color.sty in LaTeX.}%
    \renewcommand\color[2][]{}%
  }%
  \providecommand\includegraphics[2][]{%
    \GenericError{(gnuplot) \space\space\space\@spaces}{%
      Package graphicx or graphics not loaded%
    }{See the gnuplot documentation for explanation.%
    }{The gnuplot epslatex terminal needs graphicx.sty or graphics.sty.}%
    \renewcommand\includegraphics[2][]{}%
  }%
  \providecommand\rotatebox[2]{#2}%
  \@ifundefined{ifGPcolor}{%
    \newif\ifGPcolor
    \GPcolortrue
  }{}%
  \@ifundefined{ifGPblacktext}{%
    \newif\ifGPblacktext
    \GPblacktextfalse
  }{}%
  % define a \g@addto@macro without @ in the name:
  \let\gplgaddtomacro\g@addto@macro
  % define empty templates for all commands taking text:
  \gdef\gplbacktext{}%
  \gdef\gplfronttext{}%
  \makeatother
  \ifGPblacktext
    % no textcolor at all
    \def\colorrgb#1{}%
    \def\colorgray#1{}%
  \else
    % gray or color?
    \ifGPcolor
      \def\colorrgb#1{\color[rgb]{#1}}%
      \def\colorgray#1{\color[gray]{#1}}%
      \expandafter\def\csname LTw\endcsname{\color{white}}%
      \expandafter\def\csname LTb\endcsname{\color{black}}%
      \expandafter\def\csname LTa\endcsname{\color{black}}%
      \expandafter\def\csname LT0\endcsname{\color[rgb]{1,0,0}}%
      \expandafter\def\csname LT1\endcsname{\color[rgb]{0,1,0}}%
      \expandafter\def\csname LT2\endcsname{\color[rgb]{0,0,1}}%
      \expandafter\def\csname LT3\endcsname{\color[rgb]{1,0,1}}%
      \expandafter\def\csname LT4\endcsname{\color[rgb]{0,1,1}}%
      \expandafter\def\csname LT5\endcsname{\color[rgb]{1,1,0}}%
      \expandafter\def\csname LT6\endcsname{\color[rgb]{0,0,0}}%
      \expandafter\def\csname LT7\endcsname{\color[rgb]{1,0.3,0}}%
      \expandafter\def\csname LT8\endcsname{\color[rgb]{0.5,0.5,0.5}}%
    \else
      % gray
      \def\colorrgb#1{\color{black}}%
      \def\colorgray#1{\color[gray]{#1}}%
      \expandafter\def\csname LTw\endcsname{\color{white}}%
      \expandafter\def\csname LTb\endcsname{\color{black}}%
      \expandafter\def\csname LTa\endcsname{\color{black}}%
      \expandafter\def\csname LT0\endcsname{\color{black}}%
      \expandafter\def\csname LT1\endcsname{\color{black}}%
      \expandafter\def\csname LT2\endcsname{\color{black}}%
      \expandafter\def\csname LT3\endcsname{\color{black}}%
      \expandafter\def\csname LT4\endcsname{\color{black}}%
      \expandafter\def\csname LT5\endcsname{\color{black}}%
      \expandafter\def\csname LT6\endcsname{\color{black}}%
      \expandafter\def\csname LT7\endcsname{\color{black}}%
      \expandafter\def\csname LT8\endcsname{\color{black}}%
    \fi
  \fi
    \setlength{\unitlength}{0.0500bp}%
    \ifx\gptboxheight\undefined%
      \newlength{\gptboxheight}%
      \newlength{\gptboxwidth}%
      \newsavebox{\gptboxtext}%
    \fi%
    \setlength{\fboxrule}{0.5pt}%
    \setlength{\fboxsep}{1pt}%
\begin{picture}(3024.00,8784.00)%
      \csname LTb\endcsname%
      \put(1512,8564){\makebox(0,0){\strut{}}}%
    \gplgaddtomacro\gplbacktext{%
      \csname LTb\endcsname%
      \put(132,6136){\makebox(0,0)[r]{\strut{}$4$}}%
      \put(132,6470){\makebox(0,0)[r]{\strut{}$6$}}%
      \put(132,6804){\makebox(0,0)[r]{\strut{}$8$}}%
      \put(132,7138){\makebox(0,0)[r]{\strut{}$10$}}%
      \put(132,7472){\makebox(0,0)[r]{\strut{}$12$}}%
      \put(132,7806){\makebox(0,0)[r]{\strut{}$14$}}%
      \put(264,5599){\makebox(0,0){\strut{}}}%
      \put(1096,5599){\makebox(0,0){\strut{}}}%
      \put(1927,5599){\makebox(0,0){\strut{}}}%
      \put(2759,5599){\makebox(0,0){\strut{}}}%
      \put(514,6053){\makebox(0,0)[l]{\strut{}Collect by $\alpha$}}%
    }%
    \gplgaddtomacro\gplfronttext{%
      \csname LTb\endcsname%
      \put(-154,6971){\rotatebox{-270}{\makebox(0,0){\strut{}}}}%
      \put(1511,5533){\makebox(0,0){\strut{}}}%
    }%
    \gplgaddtomacro\gplbacktext{%
      \csname LTb\endcsname%
      \put(132,3501){\makebox(0,0)[r]{\strut{}$4$}}%
      \put(132,3835){\makebox(0,0)[r]{\strut{}$6$}}%
      \put(132,4169){\makebox(0,0)[r]{\strut{}$8$}}%
      \put(132,4504){\makebox(0,0)[r]{\strut{}$10$}}%
      \put(132,4838){\makebox(0,0)[r]{\strut{}$12$}}%
      \put(132,5172){\makebox(0,0)[r]{\strut{}$14$}}%
      \put(264,2964){\makebox(0,0){\strut{}}}%
      \put(1096,2964){\makebox(0,0){\strut{}}}%
      \put(1927,2964){\makebox(0,0){\strut{}}}%
      \put(2759,2964){\makebox(0,0){\strut{}}}%
      \put(514,3418){\makebox(0,0)[l]{\strut{}Collect by $y_{\text{ign}}$}}%
    }%
    \gplgaddtomacro\gplfronttext{%
      \csname LTb\endcsname%
      \put(-374,4336){\rotatebox{-270}{\makebox(0,0){\strut{}$\omega / 2 \pi$ (Hz)}}}%
      \put(1511,2898){\makebox(0,0){\strut{}}}%
    }%
    \gplgaddtomacro\gplbacktext{%
      \csname LTb\endcsname%
      \put(132,867){\makebox(0,0)[r]{\strut{}$4$}}%
      \put(132,1201){\makebox(0,0)[r]{\strut{}$6$}}%
      \put(132,1535){\makebox(0,0)[r]{\strut{}$8$}}%
      \put(132,1869){\makebox(0,0)[r]{\strut{}$10$}}%
      \put(132,2203){\makebox(0,0)[r]{\strut{}$12$}}%
      \put(132,2537){\makebox(0,0)[r]{\strut{}$14$}}%
      \put(264,330){\makebox(0,0){\strut{}$10^{1}$}}%
      \put(1096,330){\makebox(0,0){\strut{}$10^{2}$}}%
      \put(1927,330){\makebox(0,0){\strut{}$10^{3}$}}%
      \put(2759,330){\makebox(0,0){\strut{}$10^{4}$}}%
      \put(514,784){\makebox(0,0)[l]{\strut{}Collect by $T_{\text{peak}}$}}%
    }%
    \gplgaddtomacro\gplfronttext{%
      \csname LTb\endcsname%
      \put(-154,1702){\rotatebox{-270}{\makebox(0,0){\strut{}}}}%
      \put(1511,110){\makebox(0,0){\strut{}$t$ (secs)}}%
    }%
    \gplbacktext
    \put(0,0){\includegraphics{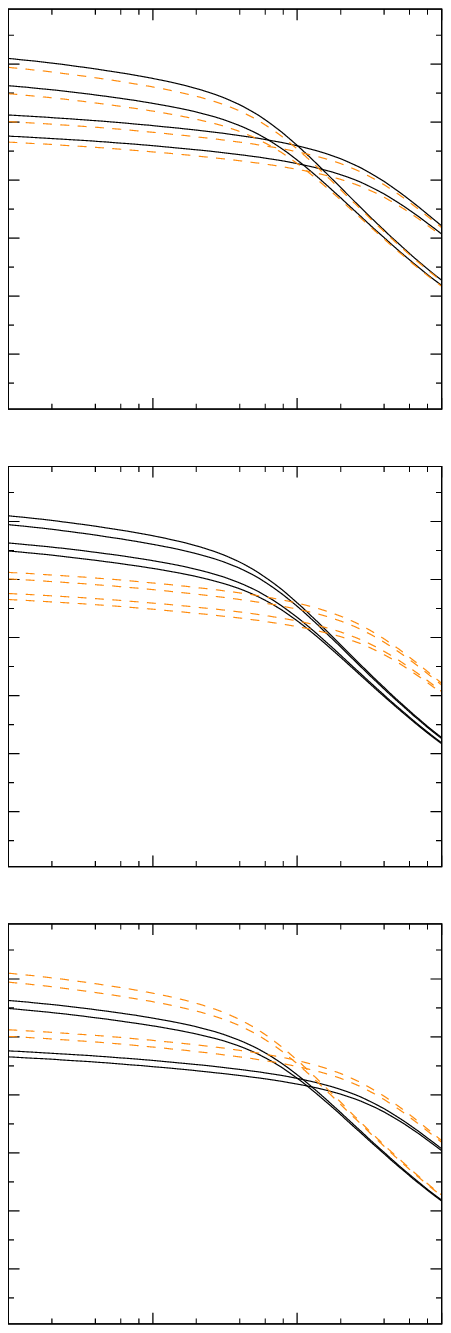}}%
    \gplfronttext
  \end{picture}%
\endgroup

%% file: images/eigenvalues-hot-crust.tex
% GNUPLOT: LaTeX picture with Postscript
\begingroup
\newcommand{\tic}[0]{\tiny} \newcommand{\ax}[0]{\normalsize}
  \makeatletter
  \providecommand\color[2][]{%
    \GenericError{(gnuplot) \space\space\space\@spaces}{%
      Package color not loaded in conjunction with
      terminal option `colourtext'%
    }{See the gnuplot documentation for explanation.%
    }{Either use 'blacktext' in gnuplot or load the package
      color.sty in LaTeX.}%
    \renewcommand\color[2][]{}%
  }%
  \providecommand\includegraphics[2][]{%
    \GenericError{(gnuplot) \space\space\space\@spaces}{%
      Package graphicx or graphics not loaded%
    }{See the gnuplot documentation for explanation.%
    }{The gnuplot epslatex terminal needs graphicx.sty or graphics.sty.}%
    \renewcommand\includegraphics[2][]{}%
  }%
  \providecommand\rotatebox[2]{#2}%
  \@ifundefined{ifGPcolor}{%
    \newif\ifGPcolor
    \GPcolortrue
  }{}%
  \@ifundefined{ifGPblacktext}{%
    \newif\ifGPblacktext
    \GPblacktextfalse
  }{}%
  % define a \g@addto@macro without @ in the name:
  \let\gplgaddtomacro\g@addto@macro
  % define empty templates for all commands taking text:
  \gdef\gplbacktext{}%
  \gdef\gplfronttext{}%
  \makeatother
  \ifGPblacktext
    % no textcolor at all
    \def\colorrgb#1{}%
    \def\colorgray#1{}%
  \else
    % gray or color?
    \ifGPcolor
      \def\colorrgb#1{\color[rgb]{#1}}%
      \def\colorgray#1{\color[gray]{#1}}%
      \expandafter\def\csname LTw\endcsname{\color{white}}%
      \expandafter\def\csname LTb\endcsname{\color{black}}%
      \expandafter\def\csname LTa\endcsname{\color{black}}%
      \expandafter\def\csname LT0\endcsname{\color[rgb]{1,0,0}}%
      \expandafter\def\csname LT1\endcsname{\color[rgb]{0,1,0}}%
      \expandafter\def\csname LT2\endcsname{\color[rgb]{0,0,1}}%
      \expandafter\def\csname LT3\endcsname{\color[rgb]{1,0,1}}%
      \expandafter\def\csname LT4\endcsname{\color[rgb]{0,1,1}}%
      \expandafter\def\csname LT5\endcsname{\color[rgb]{1,1,0}}%
      \expandafter\def\csname LT6\endcsname{\color[rgb]{0,0,0}}%
      \expandafter\def\csname LT7\endcsname{\color[rgb]{1,0.3,0}}%
      \expandafter\def\csname LT8\endcsname{\color[rgb]{0.5,0.5,0.5}}%
    \else
      % gray
      \def\colorrgb#1{\color{black}}%
      \def\colorgray#1{\color[gray]{#1}}%
      \expandafter\def\csname LTw\endcsname{\color{white}}%
      \expandafter\def\csname LTb\endcsname{\color{black}}%
      \expandafter\def\csname LTa\endcsname{\color{black}}%
      \expandafter\def\csname LT0\endcsname{\color{black}}%
      \expandafter\def\csname LT1\endcsname{\color{black}}%
      \expandafter\def\csname LT2\endcsname{\color{black}}%
      \expandafter\def\csname LT3\endcsname{\color{black}}%
      \expandafter\def\csname LT4\endcsname{\color{black}}%
      \expandafter\def\csname LT5\endcsname{\color{black}}%
      \expandafter\def\csname LT6\endcsname{\color{black}}%
      \expandafter\def\csname LT7\endcsname{\color{black}}%
      \expandafter\def\csname LT8\endcsname{\color{black}}%
    \fi
  \fi
    \setlength{\unitlength}{0.0500bp}%
    \ifx\gptboxheight\undefined%
      \newlength{\gptboxheight}%
      \newlength{\gptboxwidth}%
      \newsavebox{\gptboxtext}%
    \fi%
    \setlength{\fboxrule}{0.5pt}%
    \setlength{\fboxsep}{1pt}%
\begin{picture}(5328.00,5760.00)%
      \csname LTb\endcsname%
      \put(2664,5540){\makebox(0,0){\strut{}}}%
    \gplgaddtomacro\gplbacktext{%
      \csname LTb\endcsname%
      \put(132,3201){\makebox(0,0)[r]{\strut{}$4$}}%
      \put(132,3538){\makebox(0,0)[r]{\strut{}$6$}}%
      \put(132,3876){\makebox(0,0)[r]{\strut{}$8$}}%
      \put(132,4213){\makebox(0,0)[r]{\strut{}$10$}}%
      \put(132,4551){\makebox(0,0)[r]{\strut{}$12$}}%
      \put(132,4888){\makebox(0,0)[r]{\strut{}$14$}}%
      \put(1063,2660){\makebox(0,0){\strut{}}}%
      \put(1865,2660){\makebox(0,0){\strut{}}}%
      \put(502,3116){\makebox(0,0)[l]{\strut{}Collect by $\alpha$}}%
      \put(1865,4635){\makebox(0,0)[l]{\strut{}1 node}}%
      \put(643,3623){\makebox(0,0)[l]{\strut{}2 nodes}}%
    }%
    \gplgaddtomacro\gplfronttext{%
      \csname LTb\endcsname%
      \put(-154,4044){\rotatebox{-270}{\makebox(0,0){\strut{}}}}%
      \put(1463,2594){\makebox(0,0){\strut{}}}%
      \put(1463,5319){\makebox(0,0){\strut{}$y_{\text{ign}} = 2 \times 10^{11}$ g cm$^{-2}$}}%
    }%
    \gplgaddtomacro\gplbacktext{%
      \csname LTb\endcsname%
      \put(2664,3201){\makebox(0,0)[r]{\strut{}}}%
      \put(2664,3538){\makebox(0,0)[r]{\strut{}}}%
      \put(2664,3876){\makebox(0,0)[r]{\strut{}}}%
      \put(2664,4213){\makebox(0,0)[r]{\strut{}}}%
      \put(2664,4551){\makebox(0,0)[r]{\strut{}}}%
      \put(2664,4888){\makebox(0,0)[r]{\strut{}}}%
      \put(3595,2660){\makebox(0,0){\strut{}}}%
      \put(4397,2660){\makebox(0,0){\strut{}}}%
      \put(3034,3116){\makebox(0,0)[l]{\strut{}Collect by $\alpha$}}%
    }%
    \gplgaddtomacro\gplfronttext{%
      \csname LTb\endcsname%
      \put(2642,4044){\rotatebox{-270}{\makebox(0,0){\strut{}}}}%
      \put(3995,2594){\makebox(0,0){\strut{}}}%
      \put(3995,5319){\makebox(0,0){\strut{}$y_{\text{ign}} = 6 \times 10^{11}$ g cm$^{-2}$}}%
    }%
    \gplgaddtomacro\gplbacktext{%
      \csname LTb\endcsname%
      \put(132,761){\makebox(0,0)[r]{\strut{}$4$}}%
      \put(132,1098){\makebox(0,0)[r]{\strut{}$6$}}%
      \put(132,1436){\makebox(0,0)[r]{\strut{}$8$}}%
      \put(132,1774){\makebox(0,0)[r]{\strut{}$10$}}%
      \put(132,2112){\makebox(0,0)[r]{\strut{}$12$}}%
      \put(132,2449){\makebox(0,0)[r]{\strut{}$14$}}%
      \put(1063,220){\makebox(0,0){\strut{}$10^{3}$}}%
      \put(1865,220){\makebox(0,0){\strut{}$10^{4}$}}%
      \put(502,676){\makebox(0,0)[l]{\strut{}Collect by $T_{\text{peak}}$}}%
    }%
    \gplgaddtomacro\gplfronttext{%
      \csname LTb\endcsname%
      \put(-374,2815){\rotatebox{-270}{\makebox(0,0){\strut{}$\omega / 2 \pi$ (Hz)}}}%
      \put(2717,110){\makebox(0,0){\strut{}$t$ (secs)}}%
      \put(1463,2660){\makebox(0,0){\strut{}}}%
    }%
    \gplgaddtomacro\gplbacktext{%
      \csname LTb\endcsname%
      \put(2664,761){\makebox(0,0)[r]{\strut{}}}%
      \put(2664,1098){\makebox(0,0)[r]{\strut{}}}%
      \put(2664,1436){\makebox(0,0)[r]{\strut{}}}%
      \put(2664,1774){\makebox(0,0)[r]{\strut{}}}%
      \put(2664,2112){\makebox(0,0)[r]{\strut{}}}%
      \put(2664,2449){\makebox(0,0)[r]{\strut{}}}%
      \put(3595,220){\makebox(0,0){\strut{}$10^{3}$}}%
      \put(4397,220){\makebox(0,0){\strut{}$10^{4}$}}%
      \put(3034,676){\makebox(0,0)[l]{\strut{}Collect by $T_{\text{peak}}$}}%
    }%
    \gplgaddtomacro\gplfronttext{%
      \csname LTb\endcsname%
      \put(2642,2815){\rotatebox{-270}{\makebox(0,0){\strut{}}}}%
      \put(5249,-66){\makebox(0,0){\strut{}}}%
      \put(3995,2660){\makebox(0,0){\strut{}}}%
    }%
    \gplbacktext
    \put(0,0){\includegraphics{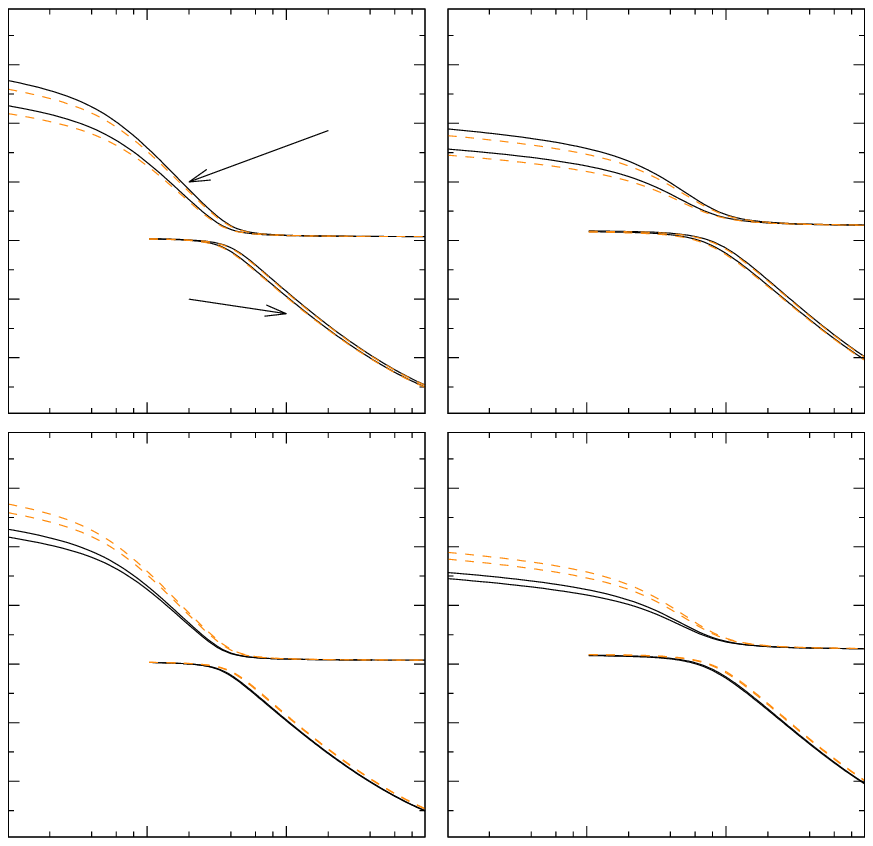}}%
    \gplfronttext
  \end{picture}%
\endgroup

%% file: images/eigenfunctions.tex
% GNUPLOT: LaTeX picture with Postscript
\begingroup
\newcommand{\tic}[0]{\tiny} \newcommand{\ax}[0]{\normalsize}
  \makeatletter
  \providecommand\color[2][]{%
    \GenericError{(gnuplot) \space\space\space\@spaces}{%
      Package color not loaded in conjunction with
      terminal option `colourtext'%
    }{See the gnuplot documentation for explanation.%
    }{Either use 'blacktext' in gnuplot or load the package
      color.sty in LaTeX.}%
    \renewcommand\color[2][]{}%
  }%
  \providecommand\includegraphics[2][]{%
    \GenericError{(gnuplot) \space\space\space\@spaces}{%
      Package graphicx or graphics not loaded%
    }{See the gnuplot documentation for explanation.%
    }{The gnuplot epslatex terminal needs graphicx.sty or graphics.sty.}%
    \renewcommand\includegraphics[2][]{}%
  }%
  \providecommand\rotatebox[2]{#2}%
  \@ifundefined{ifGPcolor}{%
    \newif\ifGPcolor
    \GPcolortrue
  }{}%
  \@ifundefined{ifGPblacktext}{%
    \newif\ifGPblacktext
    \GPblacktextfalse
  }{}%
  % define a \g@addto@macro without @ in the name:
  \let\gplgaddtomacro\g@addto@macro
  % define empty templates for all commands taking text:
  \gdef\gplbacktext{}%
  \gdef\gplfronttext{}%
  \makeatother
  \ifGPblacktext
    % no textcolor at all
    \def\colorrgb#1{}%
    \def\colorgray#1{}%
  \else
    % gray or color?
    \ifGPcolor
      \def\colorrgb#1{\color[rgb]{#1}}%
      \def\colorgray#1{\color[gray]{#1}}%
      \expandafter\def\csname LTw\endcsname{\color{white}}%
      \expandafter\def\csname LTb\endcsname{\color{black}}%
      \expandafter\def\csname LTa\endcsname{\color{black}}%
      \expandafter\def\csname LT0\endcsname{\color[rgb]{1,0,0}}%
      \expandafter\def\csname LT1\endcsname{\color[rgb]{0,1,0}}%
      \expandafter\def\csname LT2\endcsname{\color[rgb]{0,0,1}}%
      \expandafter\def\csname LT3\endcsname{\color[rgb]{1,0,1}}%
      \expandafter\def\csname LT4\endcsname{\color[rgb]{0,1,1}}%
      \expandafter\def\csname LT5\endcsname{\color[rgb]{1,1,0}}%
      \expandafter\def\csname LT6\endcsname{\color[rgb]{0,0,0}}%
      \expandafter\def\csname LT7\endcsname{\color[rgb]{1,0.3,0}}%
      \expandafter\def\csname LT8\endcsname{\color[rgb]{0.5,0.5,0.5}}%
    \else
      % gray
      \def\colorrgb#1{\color{black}}%
      \def\colorgray#1{\color[gray]{#1}}%
      \expandafter\def\csname LTw\endcsname{\color{white}}%
      \expandafter\def\csname LTb\endcsname{\color{black}}%
      \expandafter\def\csname LTa\endcsname{\color{black}}%
      \expandafter\def\csname LT0\endcsname{\color{black}}%
      \expandafter\def\csname LT1\endcsname{\color{black}}%
      \expandafter\def\csname LT2\endcsname{\color{black}}%
      \expandafter\def\csname LT3\endcsname{\color{black}}%
      \expandafter\def\csname LT4\endcsname{\color{black}}%
      \expandafter\def\csname LT5\endcsname{\color{black}}%
      \expandafter\def\csname LT6\endcsname{\color{black}}%
      \expandafter\def\csname LT7\endcsname{\color{black}}%
      \expandafter\def\csname LT8\endcsname{\color{black}}%
    \fi
  \fi
    \setlength{\unitlength}{0.0500bp}%
    \ifx\gptboxheight\undefined%
      \newlength{\gptboxheight}%
      \newlength{\gptboxwidth}%
      \newsavebox{\gptboxtext}%
    \fi%
    \setlength{\fboxrule}{0.5pt}%
    \setlength{\fboxsep}{1pt}%
\begin{picture}(4320.00,6768.00)%
      \csname LTb\endcsname%
      \put(2160,6548){\makebox(0,0){\strut{}}}%
    \gplgaddtomacro\gplbacktext{%
      \csname LTb\endcsname%
      \put(528,4745){\makebox(0,0)[r]{\strut{}$10^{-2}$}}%
      \put(528,5289){\makebox(0,0)[r]{\strut{}$10^{-1}$}}%
      \put(528,5833){\makebox(0,0)[r]{\strut{}$10^{0}$}}%
      \put(960,4145){\makebox(0,0){\strut{}}}%
      \put(1409,4145){\makebox(0,0){\strut{}}}%
      \put(1859,4145){\makebox(0,0){\strut{}}}%
    }%
    \gplgaddtomacro\gplfronttext{%
      \csname LTb\endcsname%
      \put(-44,5181){\rotatebox{-270}{\makebox(0,0){\strut{}$|\xi_x|$ (cm)}}}%
      \put(1409,6107){\makebox(0,0){\strut{}$T_{\text{crust}} = 2 \times 10^{8}$ K}}%
    }%
    \gplgaddtomacro\gplbacktext{%
      \csname LTb\endcsname%
      \put(528,2678){\makebox(0,0)[r]{\strut{}$10^{-7}$}}%
      \put(528,3308){\makebox(0,0)[r]{\strut{}$10^{-5}$}}%
      \put(528,3939){\makebox(0,0)[r]{\strut{}$10^{-3}$}}%
      \put(960,2237){\makebox(0,0){\strut{}}}%
      \put(1409,2237){\makebox(0,0){\strut{}}}%
      \put(1859,2237){\makebox(0,0){\strut{}}}%
    }%
    \gplgaddtomacro\gplfronttext{%
      \csname LTb\endcsname%
      \put(-44,3273){\rotatebox{-270}{\makebox(0,0){\strut{}$|\xi_z|$ (cm)}}}%
      \put(1409,4200){\makebox(0,0){\strut{}}}%
    }%
    \gplgaddtomacro\gplbacktext{%
      \csname LTb\endcsname%
      \put(528,915){\makebox(0,0)[r]{\strut{}$10^{24}$}}%
      \put(528,1344){\makebox(0,0)[r]{\strut{}$10^{26}$}}%
      \put(528,1773){\makebox(0,0)[r]{\strut{}$10^{28}$}}%
      \put(960,330){\makebox(0,0){\strut{}$10^{9}$}}%
      \put(1409,330){\makebox(0,0){\strut{}$10^{12}$}}%
      \put(1859,330){\makebox(0,0){\strut{}$10^{15}$}}%
    }%
    \gplgaddtomacro\gplfronttext{%
      \csname LTb\endcsname%
      \put(-44,1366){\rotatebox{-270}{\makebox(0,0){\strut{}$\mathrm{d} E / \mathrm{d} \log y$ (ergs)}}}%
      \put(1409,44){\makebox(0,0){\strut{}}}%
      \put(1409,2292){\makebox(0,0){\strut{}}}%
    }%
    \gplgaddtomacro\gplbacktext{%
      \csname LTb\endcsname%
      \put(2292,4745){\makebox(0,0)[r]{\strut{}}}%
      \put(2292,5289){\makebox(0,0)[r]{\strut{}}}%
      \put(2292,5833){\makebox(0,0)[r]{\strut{}}}%
      \put(2724,4145){\makebox(0,0){\strut{}}}%
      \put(3174,4145){\makebox(0,0){\strut{}}}%
      \put(3623,4145){\makebox(0,0){\strut{}}}%
    }%
    \gplgaddtomacro\gplfronttext{%
      \csname LTb\endcsname%
      \put(2468,5181){\rotatebox{-270}{\makebox(0,0){\strut{}}}}%
      \put(3173,4079){\makebox(0,0){\strut{}}}%
      \put(3173,6107){\makebox(0,0){\strut{}$T_{\text{crust}} = 5 \times 10^{8}$ K}}%
    }%
    \gplgaddtomacro\gplbacktext{%
      \csname LTb\endcsname%
      \put(2292,2678){\makebox(0,0)[r]{\strut{}}}%
      \put(2292,3308){\makebox(0,0)[r]{\strut{}}}%
      \put(2292,3939){\makebox(0,0)[r]{\strut{}}}%
      \put(2724,2237){\makebox(0,0){\strut{}}}%
      \put(3174,2237){\makebox(0,0){\strut{}}}%
      \put(3623,2237){\makebox(0,0){\strut{}}}%
    }%
    \gplgaddtomacro\gplfronttext{%
      \csname LTb\endcsname%
      \put(2468,3273){\rotatebox{-270}{\makebox(0,0){\strut{}}}}%
      \put(3173,2171){\makebox(0,0){\strut{}}}%
      \put(3173,4200){\makebox(0,0){\strut{}}}%
    }%
    \gplgaddtomacro\gplbacktext{%
      \csname LTb\endcsname%
      \put(2292,915){\makebox(0,0)[r]{\strut{}}}%
      \put(2292,1344){\makebox(0,0)[r]{\strut{}}}%
      \put(2292,1773){\makebox(0,0)[r]{\strut{}}}%
      \put(2724,330){\makebox(0,0){\strut{}$10^{9}$}}%
      \put(3174,330){\makebox(0,0){\strut{}$10^{12}$}}%
      \put(3623,330){\makebox(0,0){\strut{}$10^{15}$}}%
    }%
    \gplgaddtomacro\gplfronttext{%
      \csname LTb\endcsname%
      \put(2468,1366){\rotatebox{-270}{\makebox(0,0){\strut{}}}}%
      \put(2315,110){\makebox(0,0){\strut{}$y$ (g cm$^{-2})$}}%
      \put(3173,2292){\makebox(0,0){\strut{}}}%
    }%
    \gplbacktext
    \put(0,0){\includegraphics{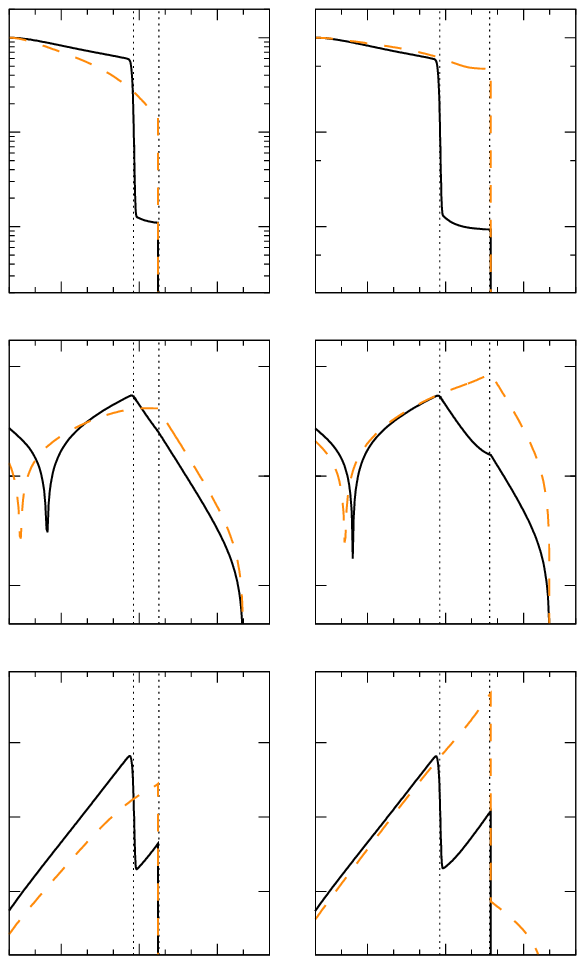}}%
    \gplfronttext
  \end{picture}%
\endgroup

%% file: images/ycrust-test.tex
% GNUPLOT: LaTeX picture with Postscript
\begingroup
\newcommand{\tic}[0]{\tiny} \newcommand{\ax}[0]{\normalsize}
  \makeatletter
  \providecommand\color[2][]{%
    \GenericError{(gnuplot) \space\space\space\@spaces}{%
      Package color not loaded in conjunction with
      terminal option `colourtext'%
    }{See the gnuplot documentation for explanation.%
    }{Either use 'blacktext' in gnuplot or load the package
      color.sty in LaTeX.}%
    \renewcommand\color[2][]{}%
  }%
  \providecommand\includegraphics[2][]{%
    \GenericError{(gnuplot) \space\space\space\@spaces}{%
      Package graphicx or graphics not loaded%
    }{See the gnuplot documentation for explanation.%
    }{The gnuplot epslatex terminal needs graphicx.sty or graphics.sty.}%
    \renewcommand\includegraphics[2][]{}%
  }%
  \providecommand\rotatebox[2]{#2}%
  \@ifundefined{ifGPcolor}{%
    \newif\ifGPcolor
    \GPcolortrue
  }{}%
  \@ifundefined{ifGPblacktext}{%
    \newif\ifGPblacktext
    \GPblacktextfalse
  }{}%
  % define a \g@addto@macro without @ in the name:
  \let\gplgaddtomacro\g@addto@macro
  % define empty templates for all commands taking text:
  \gdef\gplbacktext{}%
  \gdef\gplfronttext{}%
  \makeatother
  \ifGPblacktext
    % no textcolor at all
    \def\colorrgb#1{}%
    \def\colorgray#1{}%
  \else
    % gray or color?
    \ifGPcolor
      \def\colorrgb#1{\color[rgb]{#1}}%
      \def\colorgray#1{\color[gray]{#1}}%
      \expandafter\def\csname LTw\endcsname{\color{white}}%
      \expandafter\def\csname LTb\endcsname{\color{black}}%
      \expandafter\def\csname LTa\endcsname{\color{black}}%
      \expandafter\def\csname LT0\endcsname{\color[rgb]{1,0,0}}%
      \expandafter\def\csname LT1\endcsname{\color[rgb]{0,1,0}}%
      \expandafter\def\csname LT2\endcsname{\color[rgb]{0,0,1}}%
      \expandafter\def\csname LT3\endcsname{\color[rgb]{1,0,1}}%
      \expandafter\def\csname LT4\endcsname{\color[rgb]{0,1,1}}%
      \expandafter\def\csname LT5\endcsname{\color[rgb]{1,1,0}}%
      \expandafter\def\csname LT6\endcsname{\color[rgb]{0,0,0}}%
      \expandafter\def\csname LT7\endcsname{\color[rgb]{1,0.3,0}}%
      \expandafter\def\csname LT8\endcsname{\color[rgb]{0.5,0.5,0.5}}%
    \else
      % gray
      \def\colorrgb#1{\color{black}}%
      \def\colorgray#1{\color[gray]{#1}}%
      \expandafter\def\csname LTw\endcsname{\color{white}}%
      \expandafter\def\csname LTb\endcsname{\color{black}}%
      \expandafter\def\csname LTa\endcsname{\color{black}}%
      \expandafter\def\csname LT0\endcsname{\color{black}}%
      \expandafter\def\csname LT1\endcsname{\color{black}}%
      \expandafter\def\csname LT2\endcsname{\color{black}}%
      \expandafter\def\csname LT3\endcsname{\color{black}}%
      \expandafter\def\csname LT4\endcsname{\color{black}}%
      \expandafter\def\csname LT5\endcsname{\color{black}}%
      \expandafter\def\csname LT6\endcsname{\color{black}}%
      \expandafter\def\csname LT7\endcsname{\color{black}}%
      \expandafter\def\csname LT8\endcsname{\color{black}}%
    \fi
  \fi
    \setlength{\unitlength}{0.0500bp}%
    \ifx\gptboxheight\undefined%
      \newlength{\gptboxheight}%
      \newlength{\gptboxwidth}%
      \newsavebox{\gptboxtext}%
    \fi%
    \setlength{\fboxrule}{0.5pt}%
    \setlength{\fboxsep}{1pt}%
\begin{picture}(4320.00,5760.00)%
      \csname LTb\endcsname%
      \put(2160,5540){\makebox(0,0){\strut{}}}%
    \gplgaddtomacro\gplbacktext{%
      \csname LTb\endcsname%
      \put(792,3474){\makebox(0,0)[r]{\strut{}$11.1$}}%
      \csname LTb\endcsname%
      \put(792,4001){\makebox(0,0)[r]{\strut{}$11.2$}}%
      \csname LTb\endcsname%
      \put(792,4528){\makebox(0,0)[r]{\strut{}$11.3$}}%
      \csname LTb\endcsname%
      \put(792,5055){\makebox(0,0)[r]{\strut{}$11.4$}}%
      \csname LTb\endcsname%
      \put(1063,2990){\makebox(0,0){\strut{}}}%
      \csname LTb\endcsname%
      \put(1497,2990){\makebox(0,0){\strut{}}}%
      \csname LTb\endcsname%
      \put(2069,2990){\makebox(0,0){\strut{}}}%
      \csname LTb\endcsname%
      \put(2502,2990){\makebox(0,0){\strut{}}}%
      \csname LTb\endcsname%
      \put(2935,2990){\makebox(0,0){\strut{}}}%
      \csname LTb\endcsname%
      \put(3508,2990){\makebox(0,0){\strut{}}}%
      \csname LTb\endcsname%
      \put(3941,2990){\makebox(0,0){\strut{}}}%
    }%
    \gplgaddtomacro\gplfronttext{%
      \csname LTb\endcsname%
      \put(286,2944){\rotatebox{-270}{\makebox(0,0){\strut{}$\omega / 2 \pi$ (Hz)}}}%
      \put(2489,2704){\makebox(0,0){\strut{}}}%
      \put(2489,5429){\makebox(0,0){\strut{}$t = 10^{2}$ secs}}%
    }%
    \gplgaddtomacro\gplbacktext{%
      \csname LTb\endcsname%
      \put(792,924){\makebox(0,0)[r]{\strut{}$8$}}%
      \csname LTb\endcsname%
      \put(792,1451){\makebox(0,0)[r]{\strut{}$9$}}%
      \csname LTb\endcsname%
      \put(792,1979){\makebox(0,0)[r]{\strut{}$10$}}%
      \csname LTb\endcsname%
      \put(792,2506){\makebox(0,0)[r]{\strut{}$11$}}%
      \csname LTb\endcsname%
      \put(1063,440){\makebox(0,0){\strut{}$0.1$}}%
      \csname LTb\endcsname%
      \put(1497,440){\makebox(0,0){\strut{}$0.2$}}%
      \csname LTb\endcsname%
      \put(2069,440){\makebox(0,0){\strut{}$0.5$}}%
      \csname LTb\endcsname%
      \put(2502,440){\makebox(0,0){\strut{}$1$}}%
      \csname LTb\endcsname%
      \put(2935,440){\makebox(0,0){\strut{}$2$}}%
      \csname LTb\endcsname%
      \put(3508,440){\makebox(0,0){\strut{}$5$}}%
      \csname LTb\endcsname%
      \put(3941,440){\makebox(0,0){\strut{}$10$}}%
    }%
    \gplgaddtomacro\gplfronttext{%
      \csname LTb\endcsname%
      \put(770,395){\rotatebox{-270}{\makebox(0,0){\strut{}}}}%
      \put(2489,110){\makebox(0,0){\strut{}$y_{\text{crust}} / 10^{13}$ (g cm$^{-2}$)}}%
      \put(2489,2880){\makebox(0,0){\strut{}$t = 10^{4}$ secs}}%
      \csname LTb\endcsname%
      \put(3068,2597){\makebox(0,0)[r]{\strut{}$T_{\text{crust}} = 3 \times 10^8$ K}}%
      \csname LTb\endcsname%
      \put(3068,2377){\makebox(0,0)[r]{\strut{}$T_{\text{crust}} = 5 \times 10^8$ K}}%
    }%
    \gplbacktext
    \put(0,0){\includegraphics{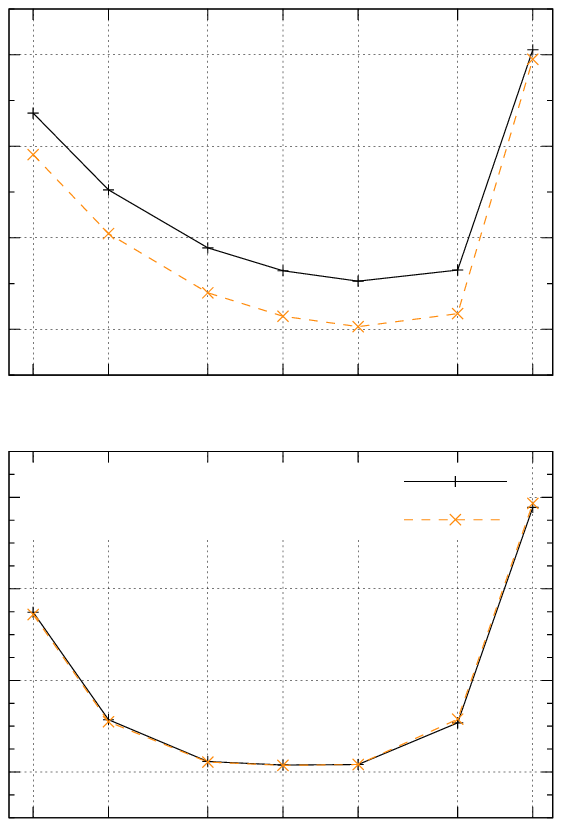}}%
    \gplfronttext
  \end{picture}%
\endgroup

%% file: images/4U-1636-536.tex
% GNUPLOT: LaTeX picture with Postscript
\begingroup
\newcommand{\tic}[0]{\tiny} \newcommand{\ax}[0]{\normalsize}
  \makeatletter
  \providecommand\color[2][]{%
    \GenericError{(gnuplot) \space\space\space\@spaces}{%
      Package color not loaded in conjunction with
      terminal option `colourtext'%
    }{See the gnuplot documentation for explanation.%
    }{Either use 'blacktext' in gnuplot or load the package
      color.sty in LaTeX.}%
    \renewcommand\color[2][]{}%
  }%
  \providecommand\includegraphics[2][]{%
    \GenericError{(gnuplot) \space\space\space\@spaces}{%
      Package graphicx or graphics not loaded%
    }{See the gnuplot documentation for explanation.%
    }{The gnuplot epslatex terminal needs graphicx.sty or graphics.sty.}%
    \renewcommand\includegraphics[2][]{}%
  }%
  \providecommand\rotatebox[2]{#2}%
  \@ifundefined{ifGPcolor}{%
    \newif\ifGPcolor
    \GPcolortrue
  }{}%
  \@ifundefined{ifGPblacktext}{%
    \newif\ifGPblacktext
    \GPblacktextfalse
  }{}%
  % define a \g@addto@macro without @ in the name:
  \let\gplgaddtomacro\g@addto@macro
  % define empty templates for all commands taking text:
  \gdef\gplbacktext{}%
  \gdef\gplfronttext{}%
  \makeatother
  \ifGPblacktext
    % no textcolor at all
    \def\colorrgb#1{}%
    \def\colorgray#1{}%
  \else
    % gray or color?
    \ifGPcolor
      \def\colorrgb#1{\color[rgb]{#1}}%
      \def\colorgray#1{\color[gray]{#1}}%
      \expandafter\def\csname LTw\endcsname{\color{white}}%
      \expandafter\def\csname LTb\endcsname{\color{black}}%
      \expandafter\def\csname LTa\endcsname{\color{black}}%
      \expandafter\def\csname LT0\endcsname{\color[rgb]{1,0,0}}%
      \expandafter\def\csname LT1\endcsname{\color[rgb]{0,1,0}}%
      \expandafter\def\csname LT2\endcsname{\color[rgb]{0,0,1}}%
      \expandafter\def\csname LT3\endcsname{\color[rgb]{1,0,1}}%
      \expandafter\def\csname LT4\endcsname{\color[rgb]{0,1,1}}%
      \expandafter\def\csname LT5\endcsname{\color[rgb]{1,1,0}}%
      \expandafter\def\csname LT6\endcsname{\color[rgb]{0,0,0}}%
      \expandafter\def\csname LT7\endcsname{\color[rgb]{1,0.3,0}}%
      \expandafter\def\csname LT8\endcsname{\color[rgb]{0.5,0.5,0.5}}%
    \else
      % gray
      \def\colorrgb#1{\color{black}}%
      \def\colorgray#1{\color[gray]{#1}}%
      \expandafter\def\csname LTw\endcsname{\color{white}}%
      \expandafter\def\csname LTb\endcsname{\color{black}}%
      \expandafter\def\csname LTa\endcsname{\color{black}}%
      \expandafter\def\csname LT0\endcsname{\color{black}}%
      \expandafter\def\csname LT1\endcsname{\color{black}}%
      \expandafter\def\csname LT2\endcsname{\color{black}}%
      \expandafter\def\csname LT3\endcsname{\color{black}}%
      \expandafter\def\csname LT4\endcsname{\color{black}}%
      \expandafter\def\csname LT5\endcsname{\color{black}}%
      \expandafter\def\csname LT6\endcsname{\color{black}}%
      \expandafter\def\csname LT7\endcsname{\color{black}}%
      \expandafter\def\csname LT8\endcsname{\color{black}}%
    \fi
  \fi
    \setlength{\unitlength}{0.0500bp}%
    \ifx\gptboxheight\undefined%
      \newlength{\gptboxheight}%
      \newlength{\gptboxwidth}%
      \newsavebox{\gptboxtext}%
    \fi%
    \setlength{\fboxrule}{0.5pt}%
    \setlength{\fboxsep}{1pt}%
\begin{picture}(4320.00,4030.00)%
    \gplgaddtomacro\gplbacktext{%
      \csname LTb\endcsname%
      \put(660,1010){\makebox(0,0)[r]{\strut{}$581.7$}}%
      \csname LTb\endcsname%
      \put(660,1622){\makebox(0,0)[r]{\strut{}$581.8$}}%
      \csname LTb\endcsname%
      \put(660,2235){\makebox(0,0)[r]{\strut{}$581.9$}}%
      \csname LTb\endcsname%
      \put(660,2847){\makebox(0,0)[r]{\strut{}$582$}}%
      \csname LTb\endcsname%
      \put(660,3459){\makebox(0,0)[r]{\strut{}$582.1$}}%
      \csname LTb\endcsname%
      \put(792,484){\makebox(0,0){\strut{}$0$}}%
      \csname LTb\endcsname%
      \put(1445,484){\makebox(0,0){\strut{}$200$}}%
      \csname LTb\endcsname%
      \put(2097,484){\makebox(0,0){\strut{}$400$}}%
      \csname LTb\endcsname%
      \put(2750,484){\makebox(0,0){\strut{}$600$}}%
      \csname LTb\endcsname%
      \put(3402,484){\makebox(0,0){\strut{}$800$}}%
      \csname LTb\endcsname%
      \put(4055,484){\makebox(0,0){\strut{}$1000$}}%
    }%
    \gplgaddtomacro\gplfronttext{%
      \csname LTb\endcsname%
      \put(22,2234){\rotatebox{-270}{\makebox(0,0){\strut{}$f = \left(\Omega - m \omega \right) / 2 \pi$ (Hz)}}}%
      \put(2423,154){\makebox(0,0){\strut{}$t - t_{\text{osc}}$ (secs)}}%
      \csname LTb\endcsname%
      \put(3068,1317){\makebox(0,0)[r]{\strut{}Observation}}%
      \csname LTb\endcsname%
      \put(3068,1097){\makebox(0,0)[r]{\strut{}$T_{\text{crust}} = 3 \times 10^8$ K}}%
      \csname LTb\endcsname%
      \put(3068,877){\makebox(0,0)[r]{\strut{}$T_{\text{crust}} = 5 \times 10^8$ K}}%
    }%
    \gplbacktext
    \put(0,0){\includegraphics{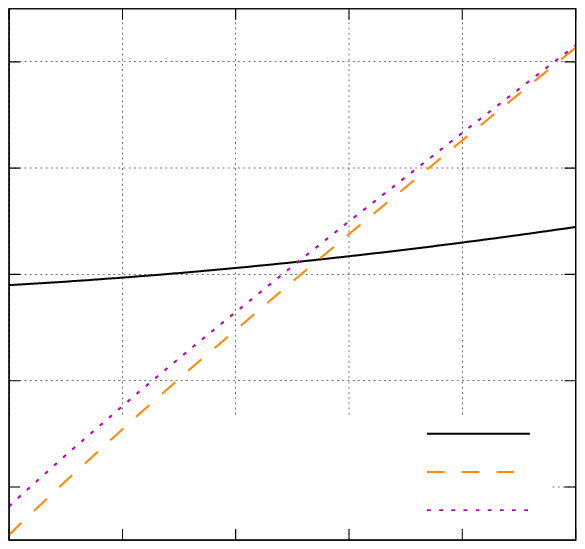}}%
    \gplfronttext
  \end{picture}%
\endgroup